\documentclass[prd,11pt,nofootinbib,showpacs,preprintnumbers,amsmath,amssymb]{revtex4}
\pdfoutput=1
\usepackage{graphicx}
\usepackage{dcolumn}

\newcommand{\cL}{{\cal L}}
\newcommand{\dvp}{{\delta \varphi}}
\newcommand{\dvpd}{\dot{\delta \varphi}}

\newcommand{\cM}{{\cal M}}

\newcommand{\bea}{\begin{eqnarray}}
\newcommand{\eea}{\end{eqnarray}}
\newcommand{\be}{\begin{equation}}
\newcommand{\ee}{\end{equation}}
\newcommand{\vp}{\varphi}

\newcommand{\pri}{{\partial_i}}
\newcommand{\prj}{{\partial_j}}
\newcommand{\half}{\frac{1}{2}}
\newcommand{\bi}{\begin{itemize}}
\newcommand{\ei}{\end{itemize}}
\newcommand{\mc}{\mathcal}

 \newcommand{\ie}{{\it i.e.~}}
 \newcommand{\eg}{{\it e.g.}}
 \newcommand{\pt}{\partial}
 \newcommand{\nn}{\nonumber}
\def\d{{\rm d}}

\newcommand{\bo}{\bar\Omega}
\newcommand{\vpbd}{\dot{\bar{\varphi}}}
\newcommand{\vpbdd}{\ddot{\bar{\varphi}}}

\begin{document}
\preprint{}
\title{A Unified Approach to Cosmic Acceleration}
\author{Minjoon Park\footnote{minjoonp@umich.edu} and Kathryn M. Zurek\footnote{kzurek@umich.edu}}
\affiliation{Michigan Center for Theoretical Physics, Ann Arbor, MI 48109}
\author{Scott Watson\footnote{gswatson@syr.edu}}
\affiliation{Michigan Center for Theoretical Physics, Ann Arbor, MI 48109 
\\and\\Department of Physics, Syracuse University, Syracuse, NY 13244}

\vspace{0.5cm}

\date{\today}

\begin{abstract}
We present a unified framework for the study of late time cosmic acceleration.  Using methods of effective field theory, we show that existing proposals for late time acceleration can be subsumed in a single framework, rather than many compartmentalized theories.  We construct the most general action consistent with symmetry principles, derive the background and perturbation evolution equations, and demonstrate that for special choices of our parameters we can reproduce results already existing in the literature.  Lastly, we lay the foundation for future work placing phenomenological constraints on the parameters of the effective theory.  Although in this paper we focus on late time acceleration, our construction also generalizes the effective field theory of inflation to the scalar-tensor and multi-field case for perturbatively constructed backgrounds.
\end{abstract}
\maketitle
\vspace{-1.2cm} 
\newpage
\tableofcontents
\newpage

\section{Introduction and Summary}
The observation of cosmic acceleration has provided a fundamental quandary for particle physics.   
The acceleration could be due to a tiny but non-zero vacuum energy density (a cosmological constant), but then the fine-tuning problem between the tiny cosmological constant and the Planck scale must be addressed, either via anthropic arguments or some new dynamics.
It is also interesting to consider, however, that the observed cosmic acceleration may be arising from new dynamics in the gravity sector.  

Experimental results suggest that at solar system scales our world is accurately described by General Relativity (GR) \cite{Damour:2008zz}, which is the unique Lorentz invariant theory of spin-2 gravitons at low energies \cite{Weinberg:1965nx}.  In order to explain the observation that the universe is accelerating on much larger scales, without invoking a cosmological constant, requires that we either add new degrees of freedom in the stress energy tensor, or that we alter the structure of general relativity (more precisely the graviton propagator).
In the former case, the new dynamics enters through an additional scalar degree of freedom, for which a new hierarchy problem between the Planck scale and the scalar mass arises that must be addressed.  In the latter case, consistent modifications to GR lead to an additional scalar degree of freedom as well.  As we will elaborate on below, one finds that the dynamics has an effective description as a scalar-tensor theory for the range of scales relevant for observations.

To date, most of the theoretical effort on dynamical dark energy models have been phenomenological in nature.  This includes Quintessence~\cite{Quintessence}, K-essence~\cite{ArmendarizPicon:2000dh}, Brans-Dicke theories~\cite{bransdicke}, modifications of the Ricci scalar ($f(R)$ models)~\cite{fofr}, first order phase transitions~\cite{Stojkovic:2007dw}, spontaneous violation of Lorentz invariance (ghost condensation)~\cite{ArkaniHamed:2003uy} and the Dvali-Gabadadze-Porrati (DGP) model~\cite{dgp}. Though on the surface, each appears to be a modification or addition to a different part of GR, all of these models can be re-written as scalar-tensor theories in a different regime.  For example, Quintessence is a scalar-tensor theory where the scalar enters the action only through its potential and interactions (and so reduces to ordinary GR plus a scalar sector), while modified gravity theories come with an additional coupling to matter in the Einstein frame (or curvature in the Jordan frame).

Thus far, the dominant approach in the literature has been to choose one model and analyze its effects on the expansion history and the matter power spectrum.  Little attempt has been made to unify these models into one fundamental, theoretical framework.  With a single framework in place, one can analyze the constraints from data on the coefficients of the terms in the effective action, and more generally determine which classes of models are consistent with the observations and more clearly see for which reasons.  Such an approach should simplify and unify the phenomenological analysis of models of dark energy.   Given the generality of our approach, we will see that many limiting cases of our action have already appeared in the literature in one form or another. 

At first it might seem hopeless to construct a single action that could account for so many different ideas to address cosmic acceleration.  However, as long as we restrict our attention phenomenologically to the long distance (low energy) regime where the theory is valid, we are able to subsume many theories with different ultraviolet behavior into one theory.  An analogous situation arises in particle physics, where many different theories for electroweak symmetry breaking in the ultraviolet describe the same low energy phenomenology of the weak and electromagnetic interactions.  To obtain the effective theory describing our low energy world, one simply writes down all the lowest order operators consistent with the symmetries of the theory.  Our goal here is to carry out this procedure to describe the low energy phenomenon of cosmic acceleration.  

In doing this, we unify all cosmic acceleration models (and modified gravity alternatives) using the approach of effective field theory (EFT).  We write down the lowest order corrections to the scalar-tensor theory and show that with these corrections all the models of dark energy described above can be reproduced.  After some work and re-writing of the lowest order terms in the action (which we discuss in more detail later), the result is remarkably simple\footnote{Throughout this paper we will work with the metric signature mostly plus ($-,+,+,+$), and with natural units $\hbar = c =1$.  For a full list of our notation and conventions we refer the reader to the appendix.}:
\bea \label{masteraction}
S &=& \int\d^4 x\sqrt{-g} \Big\{ \frac{1}{16\pi G_N}\Omega^2(\vp)R-\frac{1}{2} Z(\varphi) g^{\mu\nu}\pt_\mu\vp\pt_\nu\vp-U(\vp) + \frac{\alpha(\vp)}{\Lambda^4} (g^{\mu\nu}\pt_\mu\vp\pt_\nu\vp)^2 \Big\} + S_m . \;\;
\eea
We then proceed to do a perturbation analysis to determine cosmologically relevant parameters for the evolution of structure formation, such as the anisotropic stress.  We show that our analysis reproduces the existing results in the literature.  We leave for a second paper a more concrete application to phenomenological analysis of dark energy models, which will be useful for constraining them.

The outline of our paper is as follows.  In the next section we lay out details for arriving at the effective action (\ref{masteraction}) and make some comparisons and connections with models already existing in the literature.  In the following section, we connect (\ref{masteraction}) to observations by deriving observable quantities, such as the expansion history, the anisotropic stress and the effective Newton constant.  We also discuss how our action again reduces in special cases to existing ideas and models of dark energy. We then conclude and outline future directions for the application of our framework to data analysis and constraints on dark energy models.  To avoid obscuring the presentation, we leave a number of technical results to the appendices.  These include the full derivation of the action (\ref{masteraction}), as well as the resulting equations of motion at both the background level and for cosmological perturbations.  Our choice of conventions are summarized in Appendix \ref{appendix0}.

\section{Parametrizing Cosmic Acceleration \label{sectionII}}
As mentioned above, if we wish to construct a theory of cosmic acceleration without invoking a cosmological constant, this necessarily requires the addition of an {\em effective} scalar degree of freedom.  For the case of additional components in the stress-energy tensor this is not difficult to understand.   Whatever comprises the substance driving the expansion -- whether it be a cosmic fluid, fundamental scalar, or some other more exotic physics -- it must result in a single, dominant, adiabatic mode which respects the homogeneity and isotropy of the background.\footnote{Of course at the level of perturbations things could be more interesting.} Such a fluid can then always be written as a scalar field with (when necessary) a potential and possibly derivative interactions.\footnote{In fact, there is a more elegant way to see the appearance of this scalar as discussed in \cite{Dubovsky:2005xd} (see also \cite{ArkaniHamed:2003uy})  The scalar appearing to parametrize the physics can be thought of as the Goldstone boson of the spontaneously broken time diffeomorphism of the theory, which is simply the statement that the fluid represents a preferred frame and can be treated as a `cosmic clock' in this frame.} 
In the case of modified gravity as general relativity plus a scalar, the situation turns out to be a bit more subtle.  

\begin{figure}
\begin{center}
\includegraphics[width=5in]{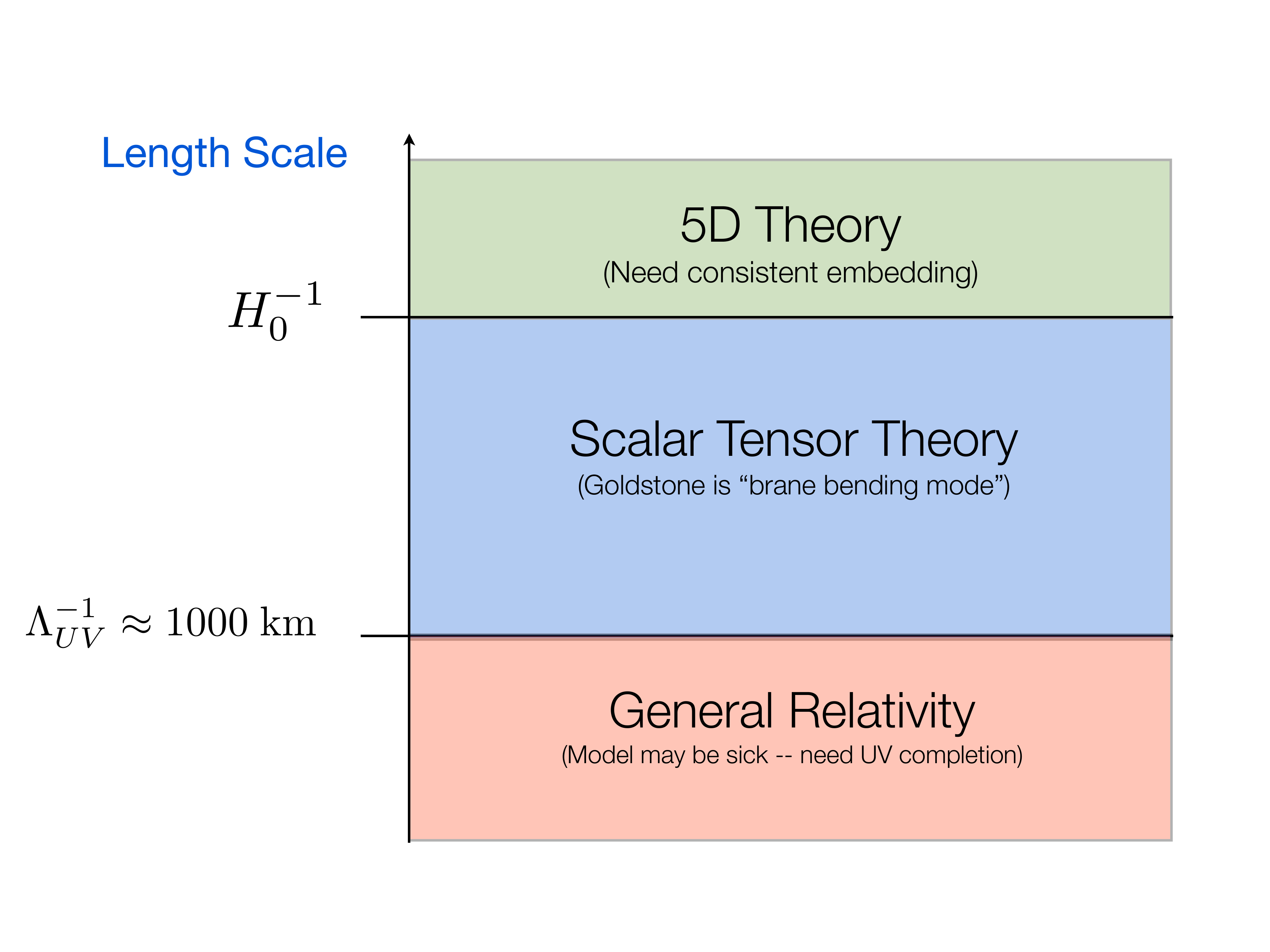}
\caption{In DGP gravity it is posited that our universe is a 3-brane embedded in a higher dimensional Minkowski space-time.  It is then argued that a weakening of gravity on large scales in our $3+1$ dimensional world could offer an alternative explanation for observations of cosmic acceleration.  For scales smaller than the cosmological horizon, but larger than around $1000$ km, one finds that the physics is described by a scalar-tensor theory.  The Goldstone boson discussed in the text is geometrically realized as the so-called  `brane bending mode'.  Although this effective description in terms of a scalar-tensor theory is valid cosmologically, for locally bound objects of higher than average density, classical non-linearities can become important before reaching the strong coupling scale $\Lambda_{UV}$ -- this is the Vainshtein effect \cite{Vainshtein:1972sx}.}
\label{fig1}
\end{center}
\end{figure}

By a `modification of gravity,' we mean that we want to alter the spin-2 structure of the graviton above solar system length scales.  This requires giving the graviton a mass, whether explicitly as in massive gravity or effectively through symmetry breaking. It is well known that  in the small wavelength limit such a mass term leads to the (in)famous van Dam-Veltman-Zakharov (vDVZ) discontinuity \cite{Zakharov:1970cc,vanDam:1970vg} as one attempts to connect smoothly to the GR limit.\footnote{This result was proven for an expansion around flat space-time. On large scales, where such an approximation would break down in the presence of a positive or negative cosmological constant the situation can change.  However in this paper we are interested in the modification of gravity as a replacement for a positive cosmological constant, so that flat space-time case is relevant.}  This presents a challenge for any modification of gravity.  In \cite{ArkaniHamed:2002sp}, it was pointed out that the physics responsible for this discontinuity can be traced to the longitudinal component of the graviton, in much the same way that occurs for massive gauge bosons in gauge theories.  The authors then demonstrated that by elevating the parameter of the broken time symmetry to a field (`Stueckelberg trick'), one can non-linearly realize the symmetry as in the analogous case of the Higgs mechanism.  This implies that below the strong coupling scale (set by the graviton mass and Newton's constant) the effective description is that of general relativity plus a derivatively coupled scalar field.  That is, in this range the modification of gravity is described by a scalar-tensor theory.  The strong coupling scale marks the scale at which the scalar-tensor effective theory breaks down and we must turn to the UV completion of the theory.  For the phenomenologically interesting case of an effective graviton mass of order the horizon scale today, this strong coupling scale implies a UV cutoff for the scalar-tensor theory of around $1/(1000 \; \mbox{km})$.  It is above this scale, that the UV completion of the theory will determine whether a consistent connection with general relativity is possible.  This Goldstone approach (where the scalar field in the scalar tensor theory is the Goldstone coming from the broken symmetry) was utilized for example in \cite{Luty:2003vm,Nicolis:2004qq} to obtain an effective field theory for DGP models, which exhibit the expected behavior as illustrated in Figure \ref{fig1}.  It has also been used to develop the effective field theory for single field inflation in \cite{Cheung:2007st}.

\subsubsection{Scalar-Tensor Effective Field Theory}
We have argued above that in order to account for cosmic acceleration in the absence of a cosmological constant requires an additional, effective scalar degree of freedom in addition to the spin-2 graviton of general relativity.  Thus, we take as our starting point the action of a scalar-tensor theory
\be\label{treeaction}
S_0 = \int\d^4 x\sqrt{-g} \left\{ \frac{m_p^2}{2}\Omega_0^2(\vp)R-\frac{1}{2} Z_0(\varphi) g^{\mu\nu}\pt_\mu\vp\pt_\nu\vp-U_0(\vp)\right\}+S_m,
\ee
where $m_p= 1/\sqrt{8 \pi G_N}=2.43 \times 10^{18}$ GeV is the reduced Planck constant, $g_{\mu \nu}$ is the space-time metric, and $S_m$ represent any matter fields that are present, which includes the standard model fields and cold dark matter.
By a conformal transformation (see Appendix \ref{appendixc}), we can move from the Jordan frame where matter is not coupled to the scalar, to the Einstein frame where the graviton (Ricci curvature) is canonically normalized. We will perform our calculations in the Jordan frame, as it is in this frame that lengths and times correspond to those measured by laboratory rulers and clocks -- \eg, measurements of supernovae will be most easily interpreted in this frame.   We stress, however, that this is only for convenience, as any conclusions drawn from experimental measurements will not depend on the choice of frame.
Because of the presence of a new scalar force mediated by $\varphi$, the effective coupling for the force measured in a Cavendish type experiment between two test masses is now given by \cite{polarski}
\be  \label{geffective}
G_{\rm eff}=G_N \Omega_0^{-2} \frac{Z_0+8m_p^2{\Omega_0^\prime}^2}{Z_0+6m_p^2{\Omega_0^\prime}^2} \,,
\ee
which only reduces to the usual Newtonian constant in the special case $\Omega_0=1$.

We are interested in constructing the most general theory of a scalar, graviton, and matter within an effective field theory approach.  This means that to the tree level action we add the leading corrections consistent with the symmetries.  Working at the level of four derivatives and after identifying terms that are equivalent up to a boundary term we find 
\bea\label{eqn:eftact}
\Delta S &=& \int\d^4 x\sqrt{-g} \Big\{ \frac{\alpha_1}{\Lambda^4} (g^{\mu\nu}\pt_\mu\vp\pt_\nu\vp)^2 + \frac{\alpha_2}{\Lambda^3} \Box\vp g^{\mu\nu}\pt_\mu\vp\pt_\nu\vp 
+ \frac{\alpha_3}{\Lambda^2} (\Box\vp)^2 \nn\\
&&\qquad+ \frac{b_1}{\Lambda^2\Lambda_m^2} T^{\mu\nu} \pt_\mu\vp\pt_\nu\vp + \frac{b_2}{\Lambda^2\Lambda_m^2} T g^{\mu\nu}\pt_\mu\vp\pt_\nu\vp 
+ \frac{b_3}{\Lambda\Lambda_m^2} T \Box\vp \nn\\
&&\qquad+ \frac{c_1}{\Lambda^2} R^{\mu\nu}\pt_\mu\vp\pt_\nu\vp + \frac{c_2}{\Lambda^2} R g^{\mu\nu}\pt_\mu\vp\pt_\nu\vp + \frac{c_3}{\Lambda} R \Box\vp \\
&&\qquad+ d_1 W^{\mu\nu\lambda\rho}W_{\mu\nu\lambda\rho} + d_2 \epsilon^{\mu\nu\lambda\rho}W_{\mu\nu}{}^{\alpha\beta}W_{\lambda\rho\alpha\beta} + d_3 R^{\mu\nu}R_{\mu\nu} + d_4 R^2 \nn\\
&&\qquad+ \frac{e_1}{\Lambda_m^4} T^{\mu\nu}T_{\mu\nu} + \frac{e_2}{\Lambda_m^4} T^2 
+ \frac{e_3}{\Lambda_m^2} R_{\mu\nu}T^{\mu\nu} + \frac{e_4}{\Lambda_m^2} RT \Big\} \,, \nn
\eea
where $\alpha_i$, $b_i$, $c_i$, $d_i$ and $e_i$ are all dimensionless functions of $\vp/\Lambda$.
We have written the matter contribution to the action (dark matter plus standard model fields) in terms of the stress energy tensor $T^{\mu}_{\nu}= {\rm diag}(-\rho, \vec{p})$ with $\rho$ and $p$ the energy density and pressure, respectively.
The Weyl tensor is
$W_{\mu \nu \rho \sigma}=R_{\mu \nu \rho \sigma}-\half \left( g_{\mu \rho} R_{\nu \sigma} - g_{\mu \sigma} R_{\nu \rho} - g_{\nu \rho} R_{\mu \sigma} + g_{\nu \sigma} R_{\mu \rho} \right) + \frac{R}{6} \left( g_{\mu \rho} g_{\nu \sigma}-g_{\nu \rho}g_{\mu \sigma}  \right)$, and $\Lambda$ and $\Lambda_m$ are the cutoffs where the scalar $\vp$ and scalar-matter effective theories break down, respectively.  The cutoff, $\Lambda_g$, for the gravity sector is implicitly assumed. The above theory becomes invalid as we approach the lowest of the three cutoffs -- we will comment more on this below.

Given the full action, $S = S_0 + \Delta S$, valid up to the four derivative level, we now must eliminate any higher derivatives appearing in the action using the equations of motion at the two derivative level, \ie those coming from $S_0$ alone.
Failure to do this can lead to spurious results, such as the appearance of ghosts in the theory coming from higher order time derivatives \cite{weinberg}.  We find when we do this, as shown in detail in Appendix \ref{appendixa}, that the most general effective action at the four derivative level is reduced to
\bea\label{theaction} 
S &=& \int\d^4 x\sqrt{-g} \Big\{ \frac{m_p^2}{2}\Omega^2(\vp)R-\frac{1}{2} Z(\varphi) g^{\mu\nu}\pt_\mu\vp\pt_\nu\vp-U(\vp) \nn\\
&&\qquad+ \frac{\alpha(\vp)}{\Lambda^4} (g^{\mu\nu}\pt_\mu\vp\pt_\mu\vp)^2 + \frac{b_1(\vp)}{\Lambda^2\Lambda_m^2} T^{\mu\nu} \pt_\mu\vp\pt_\nu\vp 
+ \frac{b_2(\vp)}{\Lambda^2\Lambda_m^2} T g^{\mu\nu}\pt_\mu\vp\pt_\nu\vp \nn\\
&&\qquad+ d_1(\vp) W^{\mu\nu\lambda\rho}W_{\mu\nu\lambda\rho} + d_2(\vp) \epsilon^{\mu\nu\lambda\rho}W_{\mu\nu}{}^{\alpha\beta}W_{\lambda\rho\alpha\beta} \\
&&\qquad+ \frac{e_1(\vp)}{\Lambda_m^4} T^{\mu\nu}T_{\mu\nu} + \frac{e_2(\vp)}{\Lambda_m^4} T^2 
+ \frac{f(\vp)}{\tilde\Lambda^4} U T \Big\} + S_m \,,\nn
\eea 
with $\tilde\Lambda$ a combination of $\Lambda$, $\Lambda_m$ and $\Lambda_g$.
All the contributions from the reduction process are absorbed into the redefinitions of $U(\varphi)$, the scalar potential and $Z$, $\alpha$, $b_i$, $e_i$ and $f$, which are all dimensionless functions of $\vp$. This is our main result, and generalizes the results for the effective theory of inflation in \cite{weinberg} to the scalar-tensor case.\footnote{This also generalizes the results of \cite{Cheung:2007st}, but to make contact with that work we must refer to our perturbative analysis appearing in the Appendices, since their EFT is done for the perturbations and not the background.  However, one naively expects these two approaches are ultimately the same, since the goldstone field $\pi$ is equivalent (non-linearly) to our scalar perturbation $\delta \varphi$ for the particular choice of comoving matter gauge.  The possible exceptions to this come from models with non-perturbative (not EFT) backgrounds, around which one studies the EFT of the fluctuations.  In those cases one would need to perform a non-linear field redefinition to demonstrate the equivalence at the level of fluctuations. Such models are of interest because of their prediction for low sound speed and a high level of non-Gaussianity, but in the approach of \cite{Cheung:2007st} this also requires us to give up the ability to determine the background which must be specified {\em a priori} from a UV complete theory.}  However, for the purpose of addressing late time cosmic acceleration this result can be simplified still further by a judicious choice of the relevant terms.  For explaining dynamics of the universe today with critical density $\rho \sim (10^{-3} $eV$)^4$ we are interested in an effective theory far below the mass of the electroweak gauge bosons and the scale of quantum gravity.  Thus, we expect that the UV cutoff of the scalar sector should be far below that where corrections to the standard model and/or quantum gravity become important, \ie $\Lambda_m,~\Lambda_g \gg \Lambda$.  
With this hierarchy among the scales, we recover our proposed action 
\bea  \label{ma2}
S &=& \int\d^4 x\sqrt{-g} \Big\{ \frac{m_p^2}{2}\Omega^2(\vp)R-\frac{1}{2} Z(\varphi) g^{\mu\nu}\pt_\mu\vp\pt_\nu\vp-U(\vp) + \frac{\alpha(\vp)}{\Lambda^4} (g^{\mu\nu}\pt_\mu\vp\pt_\nu\vp)^2 \Big\} + S_m .
\eea 

From this action there are immediately some familiar cases of dark energy proposals.
For example, ordinary Quintessence is recovered for the choice $\Omega^2=1$, $Z=1$, and $\alpha=0$.  The EFT of Quintessence is then characterized by allowing these coefficients to receive corrections, which will come with inverse powers of the cutoff.  This reproduces the results of \cite{Creminelli:2008wc} if we consider the equations of motion coming from (\ref{ma2}) linearized in perturbations and demand that the background admits a perturbative expansion (see Appendix \ref{appendixb} for the full equations).  We will discuss the correspondence with dark energy models more in the next section when discussing constraints from observations.  However, we would like to discuss the connection with modified gravity, which may not be apparent at the level of the action (\ref{ma2}).

\subsubsection{Connection with Modified Gravity}
First, let us consider $f(R)$ theories of modified gravity as effective field theories.  $f(R)$ theories appear to be a modification of the GR action which, unlike Quintessence, does not involve a light scalar degree of freedom.
The $f(R)$ action is given by 
\be \label{fr}
S_{f(R)}= \frac{m_p^2}{2} \int \d^4x \sqrt{-g}  \left\{R + m_p^2f(R) \right\}  + S_m,
\ee 
where $f(R)$ is an arbitrary function of the Ricci scalar. 
If we now instead consider the action (\ref{ma2}) with the choices $Z=0$, $\Omega^2(\varphi)=1+m_pf^\prime$, and $U(\varphi)= m_p^4/2 \left( \vp f^\prime-  f \right)$ then the variation of (\ref{ma2}) with respect to $\varphi$ gives $m_p\varphi=R$ and we recover the original action (\ref{fr}) as long as $f^\prime(\varphi) \equiv \frac{\d}{\d\varphi} f$ and $f^{''}(\varphi)$ are both non-vanishing.
Thus $f(R)$ theories are scalar-tensor theories, with the departure from GR being solely captured by the scalar field with a specific choice for a potential and a direct coupling to the Ricci scalar.  Although this class of theories can be captured by our action (\ref{ma2}), they are of little interest from an effective field theory point of view, where the generic form of the corrections would be $f(R)=c_0 + c_1 R + c_2 R^2/\Lambda + \ldots$, where $\Lambda$ is the high energy cutoff and should be the scale where classical gravity breaks down (near the Planck scale).  This is of course a UV or small wavelength modification of gravity, and does not help to address cosmological observations.  Thus any $f(R)$ theory which describes cosmic acceleration is not well described as an EFT.   On the other hand, the authors of \cite{fofr} suggested a non-local form $f(R) = \mu^4 / R$ for the correction.   Such a correction will have other theoretical problems, such as the non-unitary scattering of 
low-energy gravitons.  For these reasons, we do not consider $f(R)$ theories further, though they are in fact scalar-tensor theories.  In Appendix \ref{f(R)appendix} we discuss in detail how the poor behavior of $f(R)$ as an EFT is manifest in the scalar-tensor picture. 

As discussed above, we are more interested in modifications that, while preserving the successes of general relativity at small length scales, would alter the propagator of the spin-2 graviton at larger scales.  One proposal for such a theory is DGP gravity \cite{dgp}, where the modification comes from restricting all fields but gravity to a 3-brane embedded in five-dimensional Minkowski space-time.  The model then attempts to account for the cosmological observations of an accelerated expansion -- even in the absence of vacuum energy -- by the weakening of gravity on large scales \cite{Deffayet:2001pu}.  Although DGP in its original form is most likely ruled out as an explanation for cosmic acceleration by both experimental \cite{Schmidt:2009sg,Laszlo:2007td,Hu:2009ua} and theoretical considerations \cite{Luty:2003vm}, it still offers a valuable example of how modifications of gravity can be described and scrutinized using the methods of effective field theory. 

 The five dimensional action for the model is 
\be
S_{DGP}=2 M_5^3 \int_{\cM} \d^5x \sqrt{-G} R(G) + M_4^2 \int_{\partial \cM} \d^4x \sqrt{-g} R(g) - 4 M_5^3 \int_{\partial \cM} \d^4x \sqrt{-g} K(g),
\ee
where $G_{MN}$ is the five dimensional metric, $g_{\mu \nu}$ is our four dimensional metric restricted to the boundary or brane denoted $\partial \cM$, and $K$ is the extrinsic curvature of the brane.  The low energy (infrared) cutoff of the theory is $\Lambda_{IR} = 2M_5^3/M_4^2$, and for experiments probing energies above this scale, gravity on the brane looks effectively four dimensional.  Using DGP to try and account for the observed cosmic acceleration then suggests setting the IR cutoff at the scale of the present horizon $\Lambda^{-1}_{IR} \approx H^{-1}_0 \approx 10^{28}$ cm. Interestingly, this IR cutoff is related to the UV cutoff $\Lambda_{UV}= \left( m_p \Lambda^2_{IR} \right)^{1/3} \approx 1/(1000$ km), above which the theory becomes strongly coupled.  

Working within the regime $\Lambda_{IR} \lesssim E \lesssim \Lambda_{UV}$, it was shown\footnote{The presence of the scalar $\varphi$ was first shown for DGP in \cite{Luty:2003vm} by considering the model in the limit that $4D$ gravity decoupled from the theory, which lead to later criticism, see \eg, \cite{Gabadadze:2006tf}.  However, recently this result was shown (using different methods) to a be locally exact, and to hold even without taking a decoupling limit for the graviton \cite{galileon}.} in \cite{Luty:2003vm}  that the modification to general relativity in the DGP model can be accounted for by adding an additional scalar degree of freedom $\varphi$ -- the so-called brane bending mode -- to Einstein's theory of a massless spin two graviton. The DGP action in the Einstein frame and in the presence of matter is 
\be \label{eframedgp}
\tilde{S}_{DGP}=\int \d^4x \sqrt{-\tilde{g}} \left(\frac{m_p^2}{2} \tilde{R} - \frac{1}{2} \tilde{g}^{\mu \nu} \partial_\mu \varphi \partial_\nu \varphi - \frac{1}{\Lambda^3} \tilde{g}^{\mu \nu} \partial_\mu \varphi \partial_\nu \varphi \tilde{\Box} \varphi + \frac{1}{2m_p} \varphi \tilde{T}_{\mu}^{\mu}  \right) + \tilde{S}_m,
\ee
where $\tilde{S}_m$ is the action for any matter present, $\Lambda \equiv \Lambda_{UV}$, and the term involving the trace of the matter stress-tensor shows the non-minimal coupling in the Einstein frame (which we denote by a tilde).  
The form of the coupling to matter is a result of linearizing around a flat background, but holds in curved backgrounds if $\varphi$ is small, as is required for the theory is to be phenomenologically viable. 

Now we demonstrate that the action (\ref{eframedgp}) can be reproduced as a special case of our action (\ref{masteraction}).  To show this simply requires a bit of rearranging.  The transformation to go to the Jordan frame where matter is minimally coupled to the metric is given by
\be\label{eqn:dgpomega}
\Omega=e^{-\frac{\varphi}{\sqrt{6}m_p}},
\ee
where $\Omega$ is the conformal factor taking us from the Jordan frame to the Einstein frame, \ie $\tilde{g}_{\mu \nu} = \Omega^2 g_{\mu \nu}$.
Using the rules presented in Appendix \ref{appendixc} , we find
\be
S_{DGP}= \int \d^4x \sqrt{-g} \left(  \frac{m_p^2}{2} \Omega^2 R + L_{\rm der}  \right) + S_m, 
\ee
where $S_m$ is the minimally coupled matter action, there is no explicit kinetic term
for $\varphi$, and $L_{\rm der}$ represents derivative interactions arising from transforming the derivative interactions in (\ref{eframedgp}). Transforming the Einstein frame derivative interactions to the Jordan frame we have
\bea\label{eqn:jfdi}
-\sqrt{-\tilde{g}}\frac{1}{\Lambda^3} \tilde{g}^{\mu \nu} \partial_\mu \varphi \partial_\nu \varphi \tilde{\Box} \varphi 
&\longrightarrow& 
-\left( \Omega^4 \sqrt{-g}\right) \frac{1}{\Lambda^3} \left( \Omega^{-2} g^{\mu \nu}\right)\partial_\mu \varphi \partial_\nu \varphi  \left( \Omega^{-2} \Box \varphi + 2 \Omega^{-2} g^{\mu \nu} \Omega^{-1} \partial_\mu \Omega \partial_\nu \varphi \right) \nonumber \\
&&=\sqrt{-g} \left[ -\frac{1}{\Lambda^3} \left(\partial_\mu \varphi \right)^2 \Box \varphi +\frac{2}{\sqrt{6}m_p\Lambda^3}  \left(\partial_\mu \varphi \right)^4 \right],
\eea
where in the last step we have used the form of the conformal factor for DGP.
If we compare this with (\ref{treeaction}) and (\ref{eqn:eftact}) we see that DGP is a scalar-tensor theory with $\Omega^2=e^{-2\varphi/\sqrt{6}m_p}$, $Z=0$, and where only the first two higher derivative corrections in (\ref{eqn:eftact}) are considered with $\alpha_1=2$ and $\alpha_2=-1$ and all other terms in (\ref{eqn:eftact}) set to zero.  Thus, the simplified action (\ref{ma2}) includes the DGP model as a specific choice of parameters, but also allows for its generalization which could lead to more phenomenologically viable possibilities. It is noteworthy that because of our procedure for finding the higher derivative equations of motion in a way that is perturbatively built up from the lower order equations of motion, the resulting EFT (\ref{ma2}) will automatically be ghost free\footnote{Of course, this assumes that all coefficients are taken positive.} and no new states will appear in the theory.\footnote{A similar conclusion was reached in a very different way in \cite{galileon}, where a special symmetry (Galileon symmetry) was invoked to get a subset of operators that are also ghost free, even when one wishes to not treat the theory as an EFT.  The form of the equations for DGP in (\ref{eframedgp}) are an example of such a `special' Galileon theory. However, as an EFT, the method followed here is adequate to ensure the absence of ghosts in the regime where the EFT remains valid.} \cite{weinberg,simon}  

\subsubsection{The Effective Field Theory of Inflation}
Before moving on to observations, we would again like to emphasize that this same framework can be used for analyzing early universe acceleration, \ie inflationary models.   If we are interested in considering cosmic inflation, then we can simply decouple the matter sector (\ie send $\Lambda_m \rightarrow \infty$) and choose $\Lambda_g\approx\Lambda \approx m_p$.  In this case field redefinitions can be used to eliminate the conformal factor $\Omega^2(\varphi)$ and $Z(\varphi)$, and we reproduce the effective field theory of inflation that was presented in \cite{weinberg}.  In this paper the main conclusion was that the leading correction to the scalar sector comes from the terms of the ``DBI or K-essence type,'' a result that can be read off from (\ref{theaction}), since the term containing $\alpha(\varphi)$ survives in the limit that we decouple matter.   However, if we instead keep the matter sector (\eg, choosing it to be a another scalar field), and set $\Lambda_m \approx m_p$, this generalizes the results appearing in that paper to the two field case \cite{us}.

\section{Constraints from Observations}
In this section we connect the proposed action for parametrizing cosmic acceleration and modified gravity (\ref{ma2})
with observations to see what phenomenological constraints we can place on the parameters appearing in the fundamental Lagrangian.

\subsection{Background Evolution}

Observations of the cosmic microwave background suggest that at high redshifts ($z \approx 1100$) the universe was described to very good accuracy 
by a homogeneous and isotropic background. Thus, we consider the Jordan frame metric
\be
\d s^2=-\d t^2 + a^2(t) \left( \frac{\d r^2}{1-\kappa r^2} + r^2 \d\theta^2 + r^2 \sin^2\theta \, \d\phi^2 \right),
\ee
where $\kappa$ is the spatial curvature which can be negative, positive, or zero, and the expansion is measured in terms of the Jordan frame Hubble parameter, $H=\frac{\d}{\d t} \ln a(t)$.  In what follows we will consider the spatially flat case ($\kappa=0$), though non-trivial curvature can easily be included.  The equations of motion coming from the action (\ref{ma2}) are derived in detail in Appendix \ref{appendixa} and \ref{appendixb}.   Using (\ref{eqn:eftrmunuap}) for the choice of metric above we find the modified Einstein equations
\bea 
3 F(\varphi) \left( H^2 + H \frac{{\d}}{\d t}\ln F(\varphi) \right) &=& m_p^{-2} \left( \rho + \half Z(\varphi) \dot{\varphi}^2 + U(\varphi) + \frac{3}{\Lambda^4} \alpha(\varphi) \dot{\varphi}^4 \right), \;\;\;\; \label{eom1} \\
F(\varphi) \left( 2 \dot{H}  -H \frac{\d}{\d t} \ln F(\varphi) + \ddot{F}(\varphi)F(\varphi)^{-1} \right) &=& - m_p^{-2} \left( \rho + p +Z(\varphi) \dot{\varphi}^2 + \frac{4}{\Lambda^4} \alpha(\varphi) \dot{\varphi}^4  \right), \label{eom2}
\eea 
where we introduce $F = \Omega^2$ as the conformal factor appearing in (\ref{ma2}) to simplify the equations.  As we have argued above, these equations are general enough to give an adequate parametrization of scalar-tensor theories, a cosmological constant or dark energy ($F=1$), and modified gravity ($F \neq 1$). 
For example, if we take the conformal factor $\Omega^2=F=1$, then we recover the equations describing a universe with a perfect fluid characterized by its pressure $p$ and energy density $\rho$, in the presence of a minimally coupled scalar field $\varphi$. 
In this special case, we see that if the contribution from the fluid is negligible, then the scalar field must roll very slowly to account for a period of cosmic acceleration, \ie $\dot{\varphi} \approx 0$ if we want a de Sitter like phase $\dot{H} \approx 0$.

We would like to connect the background equations (\ref{eom1}) and (\ref{eom2}) directly with observations.   The first question to ask is -- given a particular choice of parameters appearing in these equations (corresponding to a particular choice of model to address cosmic acceleration), does such a model reproduce the observed cosmic expansion history?  To answer this question it is more convenient to cast the equations above in terms of cosmic redshift  $1+z=a_0/a$ instead of cosmic time,
\bea \label{e1}
&&3 F \left( H^2  - H^2 (1+z) \frac{\d}{\d z} \ln F \right) \nn\\
&&\quad= m_p^{-2} \left( 3H_0^2 (1+z)^3 \Omega_m^{(0)} +\frac{3}{\Lambda^4}  \alpha(\varphi) \dot{z}^4 (\partial_z\varphi)^4 + \half  Z(\varphi) \dot{z}^2 (\partial_z \varphi)^2 +U(\varphi) \right), \\
&&\partial_z^2 F +\left(  \frac{2}{1+z} + \frac{\d}{\d z} \ln H \right) \partial_z F - \left(  \frac{2}{1+z} \frac{\d}{\d z} \ln H \right) F \nonumber \\ 
&&\quad=-\frac{m_p^{-2}}{(1+z)^2H^2} \left( 3(1+z)^3 H_0^2 \Omega_m^{(0)} + Z(\varphi) \dot{z}^2 (\partial_z \varphi)^2+ \frac{4}{\Lambda^4}  \alpha(\varphi) \dot{z}^4 (\partial_z \varphi)^4 \right), \label{e2}
\eea
where we note that $\dot{z}=-(1+z)H(z)$ and we are free to set $a_0=1$.  We have assumed that radiation is negligible compared to matter, so that $\rho \propto 1/a^3$ and $p=0$, and we can express the matter density in terms of the relative abundance, \ie $\Omega_m \equiv \rho_m / \rho_c=\left( H_0/H \right)^2 \Omega_m^{(0)} (1+z)^3$, with $\rho_c=3 H^2 m_p^2$ the critical density. 

There are several ways to proceed to see if a given model is viable for reproducing the expansion history (see \eg, \cite{Sahni:2006pa}). Following the approach discussed in \cite{polarski}, and requiring a given theory to agree with the observed expansion history as derived from $\Lambda$CDM, we can obtain a constraint on the underlying parameters of the theory.  Thus a given model must reproduce the recent expansion history 
\be \label{hubblez}
H^2(z) = H_0^2 \left( \Omega_\Lambda^{(0)}+\Omega_m^{(0)} (1+z)^3 \right),
\ee
with the measured values $\Omega_\Lambda^{(0)}=0.734 \pm 0.029$, $H_0=71.0 \pm 2.5$ km/s/Mpc, and $\Omega_m=0.222 \pm 0.026$ \cite{Komatsu:2010fb}. Consider as an example Quintessence, which corresponds to setting $F=1, Z=1$ and $\alpha=0$ in (\ref{e1}) and (\ref{e2}). We can then invert these equations to find expressions for the evolution of the scalar and potential
\bea
\frac{\d\varphi}{\d z}&=&\pm \frac{m_p}{(1+z)H} \left( 2(1+z)H^2 \frac{\d}{\d z} \ln H- 3 H_0 (1+z)^3 \Omega_m \right)^{1/2}, \\
\frac{U(\varphi)}{m_p^2}&=&3H^2 -\frac{3}{2}(1+z)^3H_0^2 \Omega_m -(1+z)H^2\frac{\d}{\d z} \ln H,
\eea
which reproduces the so-called reconstruction equations found \eg~in \cite{Huterer:1998qv}. If we then use the expression (\ref{hubblez}) in these equations we see that the $\Lambda$CDM prediction can be mimicked by a specific choice of scalar field dynamics.\footnote{It should be noted that requiring a given model to reproduce the predictions of $\Lambda$CDM is actually an overly stringent requirement, since for a given set of observations at low redshift, the behavior at higher redshift can actually be much less restrictive.  Regardless, we will see that even given this more stringent constraint, it is possible for a large range of theoretical parameters to exactly reproduce the $\Lambda$CDM history.}  This is an example of the well known fact that expansion history alone cannot distinguish between different models.

As another example of a model with $F=1$, we can consider `K-essence like' models \cite{ArmendarizPicon:2000dh} where higher derivative terms are used to construct a viable model.  For these models it is standard to compare the scalar terms with those of a perfect fluid, where one finds that by defining $X \equiv -g^{\mu \nu} \partial_\mu \varphi \partial_\nu \varphi$ and noting that for $F=1$ the scalar terms in (\ref{ma2}) give the Lagrangian ${\mathcal L}=\frac{1}{2}Z(\varphi)X+\alpha(\varphi) X^2/\Lambda^4 - U(\varphi)$, the energy density and pressure are 
\bea 
\rho&=&2X \frac{\partial {\mathcal L}}{\partial X}-{\mathcal L}=\half Z(\varphi) \dot{\varphi}^2 + \frac{3}{\Lambda^4} \alpha(\varphi) \dot{\varphi}^4 + U(\varphi), \\
p&=&{\mathcal L}=\half Z(\varphi) \dot{\varphi}^2 + \frac{1}{\Lambda^4} \alpha(\varphi) \dot{\varphi}^4 - U(\varphi),
\eea
respectively. Then the condition for a viable model of cosmic acceleration can be imposed on the equation of state 
\be \label{w}
\mbox{w}=\frac{\half Z(\varphi) \dot{\varphi}^2 + \frac{1}{\Lambda^4} \alpha(\varphi) \dot{\varphi}^4 - U(\varphi)}{\half Z(\varphi) \dot{\varphi}^2 + \frac{3}{\Lambda^4} \alpha(\varphi) \dot{\varphi}^4 + U(\varphi)},
\ee
where the current bound on a constant equation of state today is $\mbox{w}_0=-0.97 \pm 0.08$ \cite{Amanullah}.

A related case is derived when the field possesses a shift symmetry $\varphi \rightarrow \varphi +c$, so that the scalar is only derivatively coupled and we have $U(\varphi)=0$, and $Z_0$ and $\alpha_0$ are both constants.  This is the particular limit of our action corresponds to that considered in ghost condensation \cite{ArkaniHamed:2003uy}.   The scalar equation of motion can be written
\be
\frac{\partial}{\partial t}  \left( a^3  \dot{\varphi} \frac{\partial {\mathcal L}(X)}{\partial X}  \right) =0,
\ee
meaning that as $a \rightarrow \infty$ either $\dot{\varphi} \rightarrow 0$ or $\partial_X {\mathcal L}(X) \rightarrow 0$.  In addition, the stability of fluctuations in such a model requires $\partial^2_X{\mathcal L}(X_0)>0$.
The case $\dot{\varphi} \rightarrow 0$ does not lead to a viable model of acceleration in the absence of a potential (as can be seen from (\ref{w})).   Considering the other case for the EFT above we find
\bea \label{gc1}
\frac{\partial {\mathcal L}}{\partial X}&=&\frac{1}{2}Z_0 + \frac{2 \alpha_0}{\Lambda^4} X=0, \\
\frac{\partial^2 {\mathcal L}}{\partial X^2}&=&\frac{2 \alpha_0}{\Lambda^4} > 0,
\label{gc2}
\eea
where in general $\alpha_0$ and $Z_0$ should be order one coefficients.  The second condition (\ref{gc2}) is in fact a general expectation derived from causality, that theories with a UV completion give positive contributions to the low energy effective action \cite{Adams:2006sv}.  As a result, cosmic acceleration and the condition (\ref{gc1}) implies the field is a ghost ($Z_0 < 0$).  Such theories stretch the validity of the EFT expansion,  since the cosmic acceleration implies $\langle \dot{\varphi}^2 \rangle \sim \Lambda^4$, unless $Z_0$ and $\alpha_0$ are less than one.  This means that as the universe begins to accelerate, the irrrelevant operators in the theory, which are normally suppressed by higher powers of $\langle \dot{\varphi}^2\rangle / \Lambda^2$, are beginning to become equally important.  This is an important caveat that must be kept in mind in applying our theory to existing dark energy models.

Another connection we would like to comment on is the relation to the EFT of Quintessence appearing recently in \cite{Creminelli:2008wc}, where the authors extended previous work using Goldstone methods to construct an EFT for inflation \cite{Cheung:2007st}.  The authors construct an effective field theory for the {\em fluctuations} around a fixed, accelerating background.  It was shown in \cite{weinberg}, for the case of inflation, that this approach yields the same final results as the methods we are taking in this paper if both the background and perturbations are treated within the EFT.  For the sake of completeness, we do a similar comparison in Appendix \ref{appendixg}, where we find that by expanding the action in quadratic fluctuations about the background values we recover the EFT of  Quintessence presented in \cite{Creminelli:2008wc}. 

We can see from (\ref{eom1}) that we can also reproduce the background expansion qualitatively as in DGP.  Recalling  $F=e^{-2 \varphi/(\sqrt{6} m_p)}$, $\alpha = 0$, $Z=0$, and that there is no potential term in DGP, we find
\be
H^2 - \frac{2 \dot{\varphi}}{\sqrt{6}m_p}H = \frac{\rho}{3 m_p^2}, 
\ee
which qualitatively reproduces the well known DGP result at leading order
\be
H^2 - \frac{H}{r_c} = \frac{\rho}{3 m_p^2},
\ee
with the identification $2 \dot{\varphi}/\sqrt{6} = m_p/r_c$.

The examples above demonstrate the ability of multiple theories to give indistinguishable results from that of a simple $\Lambda$CDM model at low red-shift.  As a more general example, combining (\ref{eom1}) and (\ref{eom2}) to eliminate $Z(\varphi)$, we find
\bea \label{redshifteom}
&&\partial^2_{z}F+ \left( \frac{\d}{\d z} \ln H - \frac{4}{1+z} \right) \partial_z F + \left( \frac{6}{(1+z)^2}-\frac{2}{1+z}\frac{\d}{\d z} \ln H \right) F
\nonumber \\
&&\quad= \frac{m_p^{-2}}{ (1+z)^2H^2} \left( 2U+\frac{2}{\Lambda^4} \alpha(\varphi) \dot{z}^4 (\partial_z{\varphi})^4
+3(1+z)^3 H_0^2  \Omega_m^{(0)} \right).
\eea
In special cases, {\em e.g.} vanishing scalar potential, by enforcing the expansion history (\ref{hubblez}) in (\ref{redshifteom}), one can solve this equation to find an $F(\varphi)$ that reproduces the observed expansion history.  The function $F$ is further constrained by fifth force experiments, fixing a boundary condition on the derivative of $F$.  This is not sufficient, however, to uniquely determine the form of $F$.  
Another approach is to have a fundamental theory that provides a specific form for the function $F$, such as DGP where $F=e^{-2\varphi/(\sqrt{6} m_p)}$.  Then (\ref{redshifteom}) provides a constraint on the model if it is to adequately reproduce the cosmic expansion history (\ref{hubblez}).  A caveat, however, is that the condition on the derivative of $F$ coming from fifth force experiments need not be enforced if a period of strong coupling for the scalar appears before reaching solar system scales.  Examples of such strong coupling behavior are exhibited by the Vainshtein effect \cite{Vainshtein:1972sx}, or the `Chameleon mechanism' \cite{Mota:2003tc,Khoury:2003aq}.

We see that whether one is looking to reconstruct a theory given the data, or whether we are interested in testing a particular model, distance measures alone are not adequate to distinguish dark energy models from modified gravity.  Instead we must combine this constraint with the growth of structure as we now discuss.

\subsection{Evolution of Density Perturbations}
We saw in the previous section that the expansion history of the background alone is not enough to distinguish different models for addressing cosmic acceleration, even though it can constrain some of the free parameters. 
This degeneracy can be confronted if we complement these considerations with probes of the growth of cosmic structure.  In this section we discuss how additional constraints can be placed on the effective theory (\ref{ma2}) by considering the growth of structure in the linear regime.

We are interested in the growth of structure coming from the evolution of density perturbations.  In Appendix \ref{appendixb} we present the equations describing the evolution of perturbations at the linear level coming from the effective theory (\ref{ma2}) in full generality.  Here, for simplicity, we will work directly in longitudinal gauge, where we fix the gauge freedom appearing in (\ref{eqn:1stvp})-(\ref{eqn:1stij}) by the choice $E=0, B=0$, so that the metric for scalar density perturbations is given by
\be
\d s^2 = -\left( 1+2 \phi \right) \d t^2 + a^2(t)  \left( 1-2 \psi \right) \delta_{ij} \d x^i \d x^j.
\ee
We will denote the background value of quantities by a bar, and so the scalar field is then $\varphi(t,\vec{x})=\bar{\varphi}(t) + \delta \varphi(t,\vec{x})$. In the matter part, we are interested in the growth of structure in the matter dominated regime where the contribution from radiation is negligible, and we have $\bar{\rho}_m \propto 1/a^3$ and $\bar{p}_m=0$.  The perturbations of the fluid are then given by $\delta \rho_m(t,\vec{x})$, $\delta p_m=0$, and $\bar{\rho} \partial_i \delta u = -\delta T_i^0$, with $\delta u$ the matter peculiar velocity potential.

It is useful to introduce the gauge invariant density contrast
\be
\delta_m=\frac{\delta \rho_m}{\bar{\rho}_m + \bar{p}_m}- 3H\delta u,
\ee
which, working in longitudinal gauge and in the period of matter domination, allows us to write the perturbed conservation equations (\ref{eqn:contii}) and (\ref{eqn:conti0}) as
\bea \label{conserv1}
\dot{\delta}_m&=&-\frac{k^2}{a^2} +3\dot{\psi} - 3 \frac{d}{dt}\left( H\delta u\right),  \\
\phi&=&-\dot{\delta u}, \label{conserv2}
\eea
where we work with the comoving wave number defined through $k^2=-\partial_i^2$, and we have assumed that matter has negligible shear (\ie $\pi=0$ in  (\ref{eqn:contii}) and (\ref{eqn:conti0})).  However, for the scalar-tensor and modified gravity cases, working in the Jordan frame we will have a non-zero anisotropic stress arising from the contribution of the non-trivial coupling $\Omega^2=F \neq 1$ appearing in (\ref{ma2}).  Indeed, 
from the non-diagonal components of (\ref{eqn:1stij}) we find
\be
\psi - \phi=  2 \frac{\Omega^\prime}{\Omega} \delta \varphi,
\ee
where we see that $\Omega \neq 1$ implies anisotropic stress, which we note would be absent in the Einstein frame where $\Omega=1$.  Note that in DGP, for example, $\Omega = e^{-\varphi/(\sqrt{6}m_p)}$, so that
$
\psi-\phi = -\frac{2}{\sqrt{6}}\frac{\delta \varphi}{m_p} \approx \frac{a^2}{k^2} \frac{\delta \rho}{3 m_p^2}
$, where (\ref{eqn:1stvp}) is used to get the last equality.
Note that this result matches onto the expressions in, \eg, \cite{Laszlo:2007td}, in the limit $\beta = 1$.  We leave for future work the exact matching in the limit $\beta \neq 1$, which we arises from super-horizon corrections, at energy scales below the infrared cut-off of the EFT.  We leave the exact matching to DGP including these super-horizon corrections for future work.
The presence of this anisotropic stress can then be measured by the fact that gravitational redshifts and weak lensing depend on the combination 
$\phi+\psi$, whereas the dynamics of non-relativistic matter depend on $\phi$ alone as we will now see.\footnote{For an overview with references see \eg~\cite{Hu:2009ua}.}

Combining the conservation equations (\ref{conserv1}) and (\ref{conserv2}), we find
\be \label{pert23}
\ddot{\delta}_m+2H \delta_m + \left(\frac{k}{a}\right)^2 \phi = 6H \frac{\d}{\d t} \left( \psi - H\delta u  \right) + 3 \frac{\d^2}{\d t^2} \left( \psi - H\delta u \right),
\ee
which is the generalization of Newtonian fluid mechanics to a cosmological background.  For observations of growth we are interested in modes which are far inside the Hubble radius, \ie $k \gg aH$ where $\lambda_p=a/k$ is the physical wavelength.  In this limit the equations  (\ref{eqn:1stvp})-(\ref{eqn:1stij})  dramatically simplify, and we find
\be
\frac{k^2}{a^2} \phi \approx - 4 \pi G_{\rm eff} \bar{\rho} \delta_m,
\ee
where $G_{\rm eff}$ is given by (\ref{geffective}). This approximation reduces to that found in \cite{Starobinsky:1998fr} for the case $\Omega=1$ and to the scalar-tensor case found in \cite{polarski}.  The fact that the small wavelength approximation agrees with the scalar-tensor case is an important check, since after using the equations of motion at the two derivative level we found that our more general Lagrangian only differs from a scalar-tensor theory by the higher derivative correction $~ \dot{\varphi}^4/\Lambda^4$, which on small time scales should be negligible.  We see that the result agrees with this expectation.  In the case of modified gravity, similar behavior has been noted for Ghost Condensation, where it was pointed out that the effect of the higher derivative correction would only now begin to become important \cite{ArkaniHamed:2003uy}.
Thus the key observable coming from this expression is the dependence of the effective Newton constant $G_{\rm eff}$ on the conformal factor $\Omega=\Omega(\varphi)$.  By using the approximate expression for the metric perturbation in (\ref{pert23}), we find the equation describing the evolution of the density contrast
\be \label{growth}
\ddot{\delta}_m+2H\dot{\delta}_m-4 \pi G_{\rm eff} \bar{\rho} \delta_m \approx 0,
\ee
which again is the result obtained in \cite{polarski}.   It is notable that this result is remarkably simple.  That is, we regain the well known equation for growth, but with the simple observation that the effective Newton constant can vary from that of the usual GR case.
Indeed, the growth of fluctuations as given by (\ref{growth}) gives an important way to distinguish various theories predicting differing values for $G_{\rm eff}$, since theories predicting larger (smaller) values of $G_{\rm eff}$ will result in structure forming faster (slower) than the standard $\Lambda$CDM scenario. 

Making use of the redshift relation $a=1/(1+z)$, we can cast this in the more 
observationally relevant form
\be
 \partial_z^2 \delta_m + \left(  2\frac{\d}{\d z} \ln H - \frac{1}{1+z}  \right) \partial_z \delta_m \approx \frac{3}{2} \left(1+z \right) \left(\frac{H_0^2}{H^2}\right) \frac{G_{\rm eff}(z)}{G_N} \Omega_m \delta_m.
\ee
\\
This equation for growth, along with the variation of the Newton constant, and the constraint (\ref{redshifteom}) coming from the expansion history, can be used to place stringent constraints on the parameters of the effective theory (\ref{masteraction}). 
Such an approach has already been used in the specific cases of DGP and $f(R)$, where it was found that these theories are not observationally viable -- see \eg~\cite{Hu:2009ua} and references within.  

\section{Conclusions and Outlook}
In this paper we have derived a formalism for combining existing proposals for dark energy and modified gravity into a single framework.  We find that the effective field theory analysis of scalar-tensor theory provides a compelling framework to accomplish such a task.  We have constructed the most general, local and unitary action, and calculated the leading corrections to fourth order in derivatives for all of the fields.  We find that we can reproduce the results of many different models of dark energy with this framework.

We have seen that the effective theory (\ref{masteraction}) can capture the crucial physics of modified gravity theories for intermediate scales between the horizon and the solar system.  However, there remain several challenges to obtaining an even more complete discussion.  It would be interesting to see if our approach could be pushed to super-horizon scales, where measurements like those of the cosmic microwave background could allow further constraints.  This has been done through a more phenomenological approach -- the so-called PPF formalism of \cite{Hu:2009ua}, however it is not clear whether the approach we have taken here can be pushed to such scales in a similar fashion.  The PPF formalism makes the crucial assumption that in models like DGP, above the horizon where the model becomes intrinsically five dimensional, if four dimensional energy/momentum conservation holds one can still proceed.  If one takes {\em all} fields into consideration, this seems a reasonable assumption, even though a given sector (like that of the scalar $\varphi$) will not be separately conserved.  We leave extending our methods to super-Horizon scales to a future publication \cite{us}.

A more challenging problem is that of small scales, as perturbations evolve into the non-linear regime.  This offers an important test of these theories, especially in the case of modified gravity, for it is in this regime that one expects the strong coupling of the scalar to reduce to GR.  Moreover, we have focused on cosmological applications, but in the presence of clustered and dense objects, we expect that something like the Vainshtein effect \cite{Vainshtein:1972sx} could become relevant and our effective theory will break down.   This breakdown is not due to the the strong coupling at the scale $\Lambda_{UV}$, but is instead a purely classical effect and could be crucial in understanding how effective field theories in the case of modified gravity can be connected to GR in the appropriate limit.

Within the regime of validity of our EFT,  below the horizon size but above the length scale of the solar system, we have argued that these theories are appropriately described by GR plus a non-minimally coupled scalar field. This fact was already well known in the case of DGP and $f(R)$ theories, but we argued here that this can be extended to all other consistent modifications as well. Within our EFT framework,  we have seen that strong observational constraints can be placed on these theories.  These constraints come from observations that restrict the expansion history and the growth of structure.  We saw that the choice of parameters in the effective theory can determine whether growth can be either faster or slower depending on the value of the effective Newton constant.  In addition, modified gravity theories result in a non-negligible anisotropic stress, which, by combining measures of structure growth with \eg~weak lensing, can be used to distinguish these models from those of dark energy.  Such an approach can be carried out in a similar way to that accomplished in the literature for the special cases of DGP and $f(R)$ (see \eg~\cite{Hu:2009ua} and references within).  We plan to continue this approach in a future publication \cite{us}, to demonstrate generally which are the features of a theory of modified gravity necessary to produce the observed acceleration and structure growth.

\acknowledgements
We would like to thank Lam Hui, Finn Larsen, and Andrew Tolley for useful discussions.
We would especially like to thank Wenjuan Fang and Dragan Huterer for useful discussions and comments on the paper.
This research was supported in part by the Department of Energy and the Michigan Center for Theoretical Physics.  The research of S.W. is also supported in part by the Michigan Society of Fellows.
S.W. would like to thank Cambridge University, DAMTP and the Mitchell Institute at Texas A\&M for hospitality.

\appendix 
\section{Notation and Conventions \label{appendix0}}
Throughout the paper, we work with metric signature ($-,+,+,+$) and  with natural units $\hbar = 1, c =1$. Our conventions are such that
\bea
\Gamma^\mu_{\nu\rho} &=& \frac{1}{2}g^{\mu\sigma}(\pt_\nu g_{\rho\sigma} + \pt_\rho g_{\nu\sigma} - \pt_\sigma g_{\nu\rho}) \,,\\
R_{\mu \nu}&=& \partial_{\lambda}\Gamma_{\mu \nu}^\lambda - \partial_\nu \Gamma_{\mu \lambda}^\lambda
+ \Gamma_{\mu \nu}^{\kappa} \Gamma^{\lambda}_{\lambda \kappa} - \Gamma_{\mu \lambda}^{\kappa} \Gamma_{\nu \kappa}^{\lambda}\,, \\
W_{\mu \nu \rho \sigma}&=&R_{\mu \nu \rho \sigma}-\half \left( g_{\mu \rho} R_{\nu \sigma} - g_{\mu \sigma} R_{\nu \rho} - g_{\nu \rho} R_{\mu \sigma} + g_{\nu \sigma} R_{\mu \rho} \right) + \frac{R}{6} \left( g_{\mu \rho} g_{\nu \sigma}-g_{\nu \rho}g_{\mu \sigma}  \right)\,, \\
\label{eqn:mset}
T_{\mu \nu} &=& -\frac{2}{\sqrt{-g}} \frac{\delta S_m}{\delta g^{\mu \nu}} ,
\eea
with $S_m$ the matter action.

\section{Effective Field Theory Analysis of a Scalar-Tensor Theory \label{appendixa}}
We start with the tree level scalar-tensor theory written in the Jordan frame:
\be\label{eqn:treeactap}
S_0 = \int\d^4 x\sqrt{-g} \Big\{ \frac{m_p^2}{2}\Omega_0^2(\vp)R-\frac{1}{2} Z_0(\varphi) g^{\mu\nu}\pt_\mu\vp\pt_\nu\vp-U_0(\vp)\Big\} + S_m \,.
\ee
Variation with respect to $g^{\mu\nu}$ gives 
\be\label{eqn:treermunua}
R_{\mu \nu} - \half g_{\mu \nu} R  =\frac{1}{m_p^2\Omega_0^2} \left[ T_{\mu\nu} + Z_0\pt_\mu\vp\pt_\nu\vp - g_{\mu\nu}\Big( \frac{1}{2} Z_0g^{\alpha\beta}\pt_\alpha\vp\pt_\beta\vp + U_0 \Big)\right]
+\frac{1}{\Omega_0^2} (\nabla_\mu\nabla_\nu - g_{\mu\nu}\Box ) \Omega_0^2 \,.
\ee
Taking the trace of this equation we find
\be\label{eqn:treer}
R =\frac{1}{m_p^2\Omega_0^2} \left[ -T+ Z_0g^{\mu\nu}\pt_\mu\vp\pt_\nu\vp + 4U_0 \right] + \frac{3}{\Omega_0^2}\Box\Omega_0^2 \,,
\ee
so that (\ref{eqn:treermunua}) in its trace-reduced form is
\be\label{eqn:treermununa}
R_{\mu\nu} = \frac{1}{m_p^2\Omega_0^2} \left[ T_{\mu\nu} - \frac{1}{2}g_{\mu\nu}T+Z_0\pt_\mu\vp\pt_\nu\vp+g_{\mu\nu}U_0\right]
+\frac{1}{\Omega_0^2}\left(\nabla_\mu\nabla_\nu+\frac{1}{2}g_{\mu\nu}\Box\right)\Omega_0^2 \,.
\ee
If we vary (\ref{eqn:treeactap}) with respect to $\vp$ we find the scalar equation of motion
\be\label{eqn:treevp}
Z_0\Box\vp+\frac{1}{2}Z_0'g^{\mu\nu}\pt_\mu\vp\pt_\nu\vp=U_0'-m_p^2\Omega_0\Omega_0'R\,.
\ee
$S_m$ gives the stress-energy tensor, (\ref{eqn:mset}), whose details become relevant later.

We add to (\ref{eqn:treeactap}) the leading EFT corrections with four derivatives:
\bea\label{eqn:eftactap}
\Delta S &=& \int\d^4 x\sqrt{-g} \Big\{ \frac{\alpha_1}{\Lambda^4} (g^{\mu\nu}\pt_\mu\vp\pt_\nu\vp)^2 + \frac{\alpha_2}{\Lambda^3} \Box\vp g^{\mu\nu}\pt_\mu\vp\pt_\nu\vp 
+ \frac{\alpha_3}{\Lambda^2} (\Box\vp)^2 \nn\\
&&\qquad+ \frac{b_1}{\Lambda^2\Lambda_m^2} T^{\mu\nu} \pt_\mu\vp\pt_\nu\vp + \frac{b_2}{\Lambda^2\Lambda_m^2} T g^{\mu\nu}\pt_\mu\vp\pt_\nu\vp 
+ \frac{b_3}{\Lambda\Lambda_m^2} T \Box\vp \nn\\
&&\qquad+ \frac{c_1}{\Lambda^2} R^{\mu\nu}\pt_\mu\vp\pt_\nu\vp + \frac{c_2}{\Lambda^2} R g^{\mu\nu}\pt_\mu\vp\pt_\nu\vp + \frac{c_3}{\Lambda} R \Box\vp \\
&&\qquad+ d_1 W^{\mu\nu\lambda\rho}W_{\mu\nu\lambda\rho} + d_2 \epsilon^{\mu\nu\lambda\rho}W_{\mu\nu}{}^{\alpha\beta}W_{\lambda\rho\alpha\beta} + d_3 R^{\mu\nu}R_{\mu\nu} + d_4 R^2 \nn\\
&&\qquad+ \frac{e_1}{\Lambda_m^4} T^{\mu\nu}T_{\mu\nu} + \frac{e_2}{\Lambda_m^4} T^2 
+ \frac{e_3}{\Lambda_m^2} R_{\mu\nu}T^{\mu\nu} + \frac{e_4}{\Lambda_m^2} RT \Big\} \,, \nn
\eea
where $\alpha_i$, $b_i$, $c_i$, $d_i$ and $e_i$ are algebraic functions of $\vp$, $W$ is the Weyl tensor
and $\Lambda$ and $\Lambda_m$ are the cutoffs for $\vp$ and the matter, respectively, and the gravitational coupling $\Lambda_g \sim m_p$ is implicit given the standard normalization of the graviton.

Next, we use the tree level (second order in derivatives) EOMs, (\ref{eqn:treermunua}-\ref{eqn:treevp}), to eliminate the higher order time derivatives in (\ref{eqn:eftactap}) as discussed in \eg, \cite{weinberg}~(see also \cite{simon}). This procedure is equivalent to solving $S_0 + \Delta S$ and then expanding the results in powers of the EFT cutoffs, and guarantees the absence of ghost degrees of freedom, which would appear if the fact that the EFT terms are \emph{small} were not taken into account. Before we apply this reduction process, note that on the right hand side of (\ref{eqn:treermunua}-\ref{eqn:treevp}) there are still double derivatives acting on the fields.
Using 
\be
\nabla_\mu\nabla_\nu X(\vp) = X'' \pt_\mu\vp\pt_\nu\vp + X'\nabla_\mu\nabla_\nu\vp\,, \quad
\Box X(\vp) = X'' g^{\mu\nu}\pt_\mu\vp\pt_\nu\vp + X' \Box\vp\,, \nn
\ee
we can solve (\ref{eqn:treer}) and (\ref{eqn:treevp}) for $R$ and $\Box\vp$ in terms of quantities with at most one derivative.
\bea\label{eqn:treerr} 
&&m_p^2\Omega_0^2\big(Z_0+6m_p^2\Omega_0'{}^2\big)R 
= -Z_0T + 4Z_0U_0 + 6m_p^2\Omega_0\Omega_0'U_0' \nn\\
&&\hspace{100pt} + \Big\{Z_0\big(Z_0+6m_p^2(\Omega_0'{}^2+\Omega_0\Omega_0'')\big)-3m_p^2\Omega_0\Omega_0'Z_0'\Big\}
g^{\mu\nu}\pt_\mu\vp\pt_\nu\vp, \, \;\; \\
\label{eqn:treevpr}
&&\big(Z_0+6m_p^2\Omega_0'{}^2\big)\Box\vp + \frac{1}{2}Z_0'g^{\mu\nu}\pt_\mu\vp\pt_\nu\vp \nn\\
&&\qquad = U_0' - \frac{\Omega_0'}{\Omega_0} \big\{ 4U_0 - T + \big(Z_0+6m_p^2(\Omega_0'{}^2+\Omega_0\Omega_0'')\big) g^{\mu\nu}\pt_\mu\vp\pt_\nu\vp \big\}. \, \;\;
\eea
We can rewrite the right hand side of (\ref{eqn:treermunua}) to get 
\bea\label{eqn:treermunur}
m_p^2\Omega_0^2 R_{\mu\nu} &=& T_{\mu\nu} + \big(Z_0+2m_p^2(\Omega_0'{}^2+\Omega_0\Omega_0'')\big)\pt_\mu\vp\pt_\mu\vp
+2m_p^2\Omega_0\Omega_0'\nabla_\mu\nabla_\nu\vp \nn\\
&&+ g_{\mu\nu}\big(Z_0+6m_p^2\Omega_0'{}^2\big)^{-1} \Big\{ - \big( \frac{Z_0}{2}+2m_p^2\Omega_0'{}^2 \big) T
+ \big( Z_0+2m_p^2\Omega_0'{}^2\big)U_0 + m_p^2\Omega_0\Omega_0' U_0' \nn\\
&&\hspace{95pt} + m_p^2\Omega_0\big(\Omega_0'' Z_0-\frac{1}{2}\Omega_0'Z_0'\big) g^{\mu\nu}\pt_\mu\vp\pt_\nu\vp \Big\} \,.
\eea
Now let us perform the reduction of (\ref{eqn:eftactap}): 
\bi
\item $\alpha_1$, $b_1$, $b_2$, $d_1$, $d_2$, $e_1$ and $e_2$ terms do not have two time derivatives and do not need to be reduced. 
\item With (\ref{eqn:treevpr}), $\alpha_2$ term is made to contribute to $Z_0$, $\alpha_1$ and $b_2$.
\item $\alpha_3$ contributes to $Z_0$, $U_0$, $\alpha_1$, $b_2$, $e_2$, and it also generates a term like $f(\vp) U_0T$.
\item $b_3$ contributes to $b_2$, $e_2$ and $f(\vp) U_0T$.
\item $c_1$ is nontrivial due to $\nabla_\mu\nabla_\nu\vp$ in (\ref{eqn:treermunur}). But since 
\bea
&&\sqrt{-g}\,f(\vp) \nabla_\mu\nabla_\nu\vp \nabla^\mu\vp \nabla^\nu\vp = \frac{1}{2}\sqrt{-g}\,f\nabla^\mu\vp \nabla_\mu(\nabla^\mu\vp\pt_\mu\vp) \nn\\
&&\qquad= -\frac{1}{2}\sqrt{-g}\,\{ f'(\nabla^\mu\vp\pt_\mu\vp)^2 + f \Box\vp \nabla^\mu\vp\pt_\mu\vp \}\,, \nn
\eea
it reduces itself to $\alpha_1$ and $\alpha_2$ types. Therefore, $c_1$ contributes to $Z_0$, $\alpha_1$, $b_1$ and $b_2$.
\item $c_2$: $Z_0$, $\alpha_1$ and $b_2$.
\item $c_3$: $Z_0$, $U_0$, $\alpha_1$, $b_2$, $e_2$ and $f(\vp) U_0T$.
\item $d_3$: $Z_0$, $U_0$, $\alpha_1$, $b_1$, $b_2$, $e_1$, $e_2$ and $f(\vp) U_0T$.
\item $d_4$: $Z_0$, $U_0$, $\alpha_1$, $b_2$, $e_2$ and $f(\vp) U_0T$.
\item $e_3$: $b_1$, $b_2$, $e_1$, $e_2$ and $f(\vp) U_0T$.
\item $e_4$: $b_2$, $e_2$ and $f(\vp) U_0T$.
\ei

In summary, the EFT corrected action has the form of
{\small
\bea\label{eqn:act}
S &=& \int\d^4 x\sqrt{-g} \Big\{ \frac{m_p^2}{2}\Omega^2(\vp)R-\frac{1}{2} Z(\varphi) g^{\mu\nu}\pt_\mu\vp\pt_\nu\vp-U(\vp) \nn\\
&&\qquad+ \frac{\alpha(\vp)}{\Lambda^4} (g^{\mu\nu}\pt_\mu\vp\pt_\mu\vp)^2 + \frac{b_1(\vp)}{\Lambda^2\Lambda_m^2} T^{\mu\nu} \pt_\mu\vp\pt_\nu\vp 
+ \frac{b_2(\vp)}{\Lambda^2\Lambda_m^2} T g^{\mu\nu}\pt_\mu\vp\pt_\nu\vp \nn\\
&&\qquad+ d_1 W^{\mu\nu\lambda\rho}W_{\mu\nu\lambda\rho} + d_2 \epsilon^{\mu\nu\lambda\rho}W_{\mu\nu}{}^{\alpha\beta}W_{\lambda\rho\alpha\beta} 
+ \frac{e_1}{\Lambda_m^4} T^{\mu\nu}T_{\mu\nu} + \frac{e_2}{\Lambda_m^4} T^2 + \frac{f(\vp)}{\tilde\Lambda^4} U T \Big\} + S_m \,,
\eea}
where $\tilde\Lambda^4 \sim \tilde M^4(1+m_p^2|\Omega'|^2)$ with $\tilde M^4=\Lambda^4$, $\Lambda^2\Lambda_m^2$, $\Lambda^2m_p^2$, $\Lambda_m^2m_p^2$ or $m_p^4$.
All the contributions from the reduction process are absorbed into the redefinition of $Z$, $U$, $\alpha$, $b_i$, $e_i$ and $f$. 

So far, we have not assumed any hierarchy among $\Lambda$, $\Lambda_m$ and $m_p$. 
In the early Universe where $\Lambda$ may not be much different from $m_p$, all the EFT corrections can be of a similar size and 
we have to take all of them into consideration.
However, for the purpose of addressing late time cosmic acceleration this result can be simplified still further by a judicious choice of the relevant terms.  For explaining dynamics of the universe today with critical density $\rho \sim (10^{-3} $eV$)^4$ we are interested in an effective theory far below the mass of the electroweak gauge bosons and the scale of quantum gravity.  Thus, we expect that the UV cutoff of the scalar sector should be far below that where corrections to the standard model and/or quantum gravity become important, \ie $\Lambda_m,~\Lambda_g \gg \Lambda$.  
Thus, $b$, $d$, $e$ and $f$ terms can be neglected compared to $\alpha$.
Then, the EFT corrected action for the dark energy analysis is
\bea\label{eqn:deact}
S = \int\d^4 x\sqrt{-g} \Big\{ \frac{m_p^2}{2}\Omega^2(\vp)R-\frac{1}{2} Z(\varphi) g^{\mu\nu}\pt_\mu\vp\pt_\nu\vp-U(\vp) 
+ \frac{\alpha(\vp)}{\Lambda^4} (g^{\mu\nu}\pt_\mu\vp\pt_\nu\vp)^2 \Big\} + S_m \,,
\eea
and the corresponding EOMs are
\bea\label{eqn:eftrmunuap}
m_p^2\Omega^2R^\mu_\nu &=& T^\mu_\nu - \frac{1}{2}\delta^\mu_\nu T+Zg^{\mu\rho}\pt_\rho\vp\pt_\nu\vp+\delta^\mu_\nu U
+m_p^2\big(\nabla^\mu\nabla_\nu+\frac{1}{2}\delta^\mu_\nu\Box\big)\Omega^2 \nn\\
&&- \frac{\alpha}{\Lambda^4}(g^{\rho\sigma}\pt_\rho\vp\pt_\sigma\vp)^2\big( 4g^{\mu\kappa}\pt_\kappa\vp\pt_\nu\vp - \delta^\mu_\nu g^{\kappa\lambda}\pt_\kappa\vp\pt_\lambda\vp \big) \,, \\
\label{eqn:eftvpap}
Z\Box\vp &=& U'-m_p^2\Omega\Omega'R - \frac{1}{2}Z'g^{\mu\nu}\pt_\mu\vp\pt_\nu\vp \nn\\
&&+ \frac{3\alpha'}{\Lambda^4}(g^{\mu\nu}\pt_\mu\vp\pt_\nu\vp)^2
+ \frac{4\alpha}{\Lambda^4} ( \Box\vp g^{\mu\nu}\pt_\mu\vp\pt_\mu\vp+2\pt_\mu\vp\pt_\nu\vp\nabla^\mu\nabla^\nu\vp ) \,.
\eea

\section{Cosmological Perturbations}\label{appendixb}
In this section, we obtain the perturbative expansion of the full EOMs, (\ref{eqn:eftrmunuap}-\ref{eqn:eftvpap}). 
For cosmological perturbations we work with coordinate time, assume negligible spatial curvature (this can easily be included), and parametrize the metric and scalar perturbations as
\bea \label{pertmetric}
\d s^2 &=& - \left[1+2 \phi(t,\vec{x}) \right] \d t^2 + 2 a(t) \partial_i B(t,\vec{x}) \d t \d x^i \nn\\
&&\qquad\qquad + a(t)^2 \big[ \left(1-2 \psi(t,\vec{x}) \right) \delta_{ij} +2 \partial_i \partial_j E(t,\vec{x}) \big]  \d x^i \d x^j \,, \\
\vp(t,\vec{x}) &=& \bar\vp(t) + \delta\vp(t,\vec{x}),
\label{perturbations}
\eea
respectively, where bars indicate background quantities. Matter perturbations are given by
{\small\bea
T^0_0 &=& -\bar \rho - \delta\rho \,,\;\; T^0_i = (\bar\rho+\bar p)\pt_i\delta u \,,\;\;T^i_j = \delta_{ij} \bar p + \delta_{ij}\delta p+\pt_i\pt_j\pi \,,\;\; T = -\bar\rho+3\bar p - \delta\rho+3\delta p+\pt_i^2\pi \,.
\eea}
The matter satisfies the conservation equation, $\nabla_\mu T^\mu_\nu = 0$, whose zeroth order piece is
{\small\be
\dot{\bar\rho} + 3H (\bar\rho+\bar p) = 0\,,
\ee}
and first order parts are
{\small\bea\label{eqn:contii}
 \pt_i \Big[ \delta p+\pt_i^2\pi+\pt_t[(\bar\rho+\bar p)\delta u]+3H (\bar\rho+\bar p)\delta u + (\bar\rho+\bar p)\phi \Big]&=&0 \,,\\
\label{eqn:conti0}
 \dot{\delta\rho} + 3H(\delta\rho+\delta p)+\pt_i^2\Big( -\frac{\bar\rho+\bar p}{a}B + \frac{\bar\rho+\bar p}{a^2}\delta u + H\pi \Big)
+(\bar\rho+\bar p)(-3\dot\psi+\pt_i^2\dot E)&=&0\,.
\eea}

The zeroth order EOMs obtained from (\ref{eqn:eftrmunuap}-\ref{eqn:eftvpap}) are
{\small\bea\label{eqn:0thvp}
\vp&:& \;  \bar U' + \frac{\bo'}{\bo}(-\bar\rho+3\bar p-4\bar U)+\Big(3H\vpbd + \vpbd^2\frac{\bo'}{\bo} \Big)(\bar Z+6\bo'{}^2)
+ \frac{1}{2\vpbd} \pt_t[\vpbd^2 (\bar Z+6\bo'{}^2)] \nn\\
&&+ 12H\vpbd^3 \frac{\bar\alpha}{\Lambda^4} + \frac{3}{\vpbd} \pt_t[\vpbd^4 \frac{\bar\alpha}{\Lambda^4}] =0\,, \\
\label{eqn:0th00}
00&:& \;  \frac{\bar\rho+3\bar p}{2} -\bar U + 3(H^2+\dot H)\bo^2+\vpbd^2\bar Z+3(H\vpbd + \vpbdd)\bo\bo'+3\vpbd^2(\bo'{}^2+\bo\bo'')
+3\vpbd^4\frac{\bar\alpha}{\Lambda^4}=0 \,, \\
\label{eqn:0thii}
ii&:& \;  \frac{-\bar\rho+\bar p}{2} - \bar U + (3H^2+\dot H)\bo^2 +(5H\vpbd + \vpbdd)\bo\bo'+\vpbd^2(\bo'{}^2+\bo\bo'')
-\vpbd^4\frac{\bar\alpha}{\Lambda^4}=0 \,,
\eea}
where $\bar F' = \frac{\d F(\vp)}{\d\vp}\Big|_{\bar\vp}$, $\bo = m_p\Omega(\bar\vp)$ and an overdot implies
a time derivative. They determine the expansion history of a model, or we can use them to constrain a model to give the observed expansion history of the universe.

The first order EOMs, slightly simplified by using (\ref{eqn:0thvp}-\ref{eqn:0thii}), are
{\small\bea\label{eqn:1stvp}
\vp&:& \; \frac{\bo'}{\bo}(-\delta\rho+3\delta p+\pt_i^2\pi) =  \Big(\bar Z+6\bo'{}^2 + 4\vpbd^2\frac{\bar\alpha}{\Lambda^4}\Big)\Big\{ \frac{\pt_i^2\dvp}{a^2} +3\vpbd\dot\psi+\vpbd\big(\partial_i^2\frac{B}{a}-\partial_i^2\dot E\big) \Big\} \nn\\
&&\qquad- \Big(\bar Z+6\bo'{}^2 + 12\vpbd^2\frac{\bar\alpha}{\Lambda^4}\Big) (\ddot\dvp-\vpbd\dot\phi) \nn\\
&&\qquad+ \Big\{ 2\bar U' + \frac{2\bo'}{\bo}(-\bar\rho+3\bar p-4\bar U) + (3H\vpbd+2\vpbdd)(\bar Z+6\bo'{}^2) 
-12H\vpbd^3\frac{\bar\alpha}{\Lambda^4}-6\vpbd^4\frac{\bar\alpha'}{\Lambda^4} \Big\} \frac{\dot\dvp}{\vpbd} \\
&&\qquad+ \Big\{ \frac{\bo'}{\bo}(-\dot{\bar\rho}+3\dot{\bar p}) + \Big( 3\pt_t[H\vpbd]+2\vpbd\vpbdd\frac{\bo'}{\bo} \Big)(\bar Z+6\bo'{}^2)+\pt_t[\vpbdd(\bar Z+6\bo'{}^2)]
+ 12\pt_t[H\vpbd^3] \frac{\bar\alpha}{\Lambda^4} + 12 \pt_t[\vpbd^2\vpbdd\frac{\bar\alpha}{\Lambda^4}]\Big\} \frac{\dvp}{\vpbd} \nn\\
&&\qquad- 2\Big\{ \bar U'+\frac{\bo'}{\bo}(-\bar\rho+3\bar p-4\bar U) - 12H\vpbd^3\frac{\bar\alpha}{\Lambda^4} 
- \frac{3}{\vpbd}\pt_t[\vpbd^4\frac{\bar\alpha}{\Lambda^4}] \Big\} \phi \,, \nn\\
\label{eqn:1st00}
00&:& \; -\frac{\delta\rho+3\delta p+\pt_i^2\pi}{2} = 2\vpbd\bar Z(\dot\dvp-\vpbd\phi)+(\vpbd^2\bar Z'-\bar U')\dvp \nn\\
&&\qquad-\bo^2\Big\{ \frac{\partial_i^2\phi}{a^2}+3H\dot\phi+6(H^2+\dot H)\phi+3\ddot\psi+6H\dot\psi+\partial_i^2\frac{\dot B}{a}-\partial_i^2\ddot E
+H\big(\partial_i^2\frac{B}{a}-2\partial_i^2\dot E\big) \Big\} \\
&&\qquad+\bo\bo'\Big\{ -\frac{\partial_i^2\dvp}{a^2}+3\ddot\dvp+3H\dot\dvp+6(H^2+\dot H)\dvp-3\vpbd(\dot\phi+\dot\psi)-6(H\vpbd+\vpbdd)\phi 
-\vpbd\big(\partial_i^2\frac{B}{a}-\partial_i^2\dot E\big)\Big\} \nn\\
&&\qquad+3(\bo'{}^2+\bo\bo'') \{ 2\vpbd\dot\dvp+(H\vpbd+\vpbdd)\dvp-2\vpbd^2\phi \}
+3\vpbd^2(3\bo'\bo''+\bo\bo''')\dvp \nn\\
&&\qquad+12\vpbd^3\frac{\bar\alpha}{\Lambda^4}(\dot\dvp-\vpbd\phi)+3\vpbd^4\frac{\bar\alpha'}{\Lambda^4}\dvp \,, \nn\\
\label{eqn:1st0i}
0i&:& \; (\bar\rho+\bar p)\pt_i\delta u = \pri \Big[ \bo\bo'\Big(2\dot\dvp-H\dvp-2\vpbd\phi\Big)-2\bo^2(H\phi+\dot\psi) 
+\Big( \vpbd\bar Z + 2\vpbd\bo'{}^2+2\vpbd\bo\bo''+4\vpbd^3\frac{\bar\alpha}{\Lambda^4}\Big)\dvp \Big] \,, \\
\label{eqn:1stij} 
ij&:& \; \pt_i\pt_j\pi-\frac{1}{2}\delta_{ij}(-\delta\rho+\delta p +\pt_k^2\pi) \nn\\
&&\quad= \pri\prj\Big[ \bo^2\Big( \frac{-\phi+\psi}{a^2}-H\big(2\frac{B}{a}-3\dot E\big)-\frac{\dot B}{a}+\ddot E \Big) 
+ \bo\bo'\Big( -\frac{2}{a^2}\dvp-2\vpbd\big(\frac{B}{a}-\dot E\big) \Big) \Big] \nn\\
&&\qquad+ \delta_{ij}\Big[ - \bar U'\dvp + \bo^2 \Big( -H\dot\phi-2(\dot H+3H^2)\phi+\frac{\partial_k^2\psi}{a^2}-\ddot\psi-6H\dot\psi
-H\big(\partial_k^2\frac{B}{a}-\partial_k^2\dot E\big) \Big) \\
&&\qquad+\bo\bo'\Big( -\frac{\partial_k^2\dvp}{a^2}+\ddot\dvp+5H\dot\dvp+2(3H^2+\dot H)\dvp-\vpbd\dot\phi-2(5H\vpbd+\vpbdd)\phi
-5\vpbd\dot\psi-\vpbd\big(\partial_k^2\frac{B}{a}-\partial_k^2\dot E\big) \Big) \nn\\
&&\qquad+(\bo'{}^2+\bo\bo'')\{2\vpbd\dot\dvp+(5H\vpbd+\vpbdd)\dvp-2\vpbd^2\phi \} 
+\vpbd^2(3\bo'\bo''+\bo\bo''')\dvp \nn\\
&&\qquad+4\vpbd^3\frac{\bar\alpha}{\Lambda^4}(-\dot\dvp+\vpbd\phi)-\vpbd^4\frac{\bar\alpha'}{\Lambda^4}\dvp \Big] \,. \nn
\eea}
These EOMs can be used to further specify or constrain a model. By solving (\ref{eqn:1st00}) and the trace of (\ref{eqn:1stij}) for $\pt_i^2\phi$ and $\pt_i^2\psi$ and using the zeroth order EOMs together with 
(\ref{eqn:1st0i}) and (\ref{eqn:contii}), we can get the Poisson equations for the gravitational potentials:
{\small\bea\label{eqn:boxpsi}
\pt_i^2\psi &=& \frac{a^2}{2\bo^2}\delta\rho - \frac{3a^2}{2\bo^2}H(\bar\rho+\bar p)\delta u 
+3a^2\vpbd\frac{\bo'}{\bo}\dot\psi - \frac{a^2}{2\bo^2}\big(\vpbd^2\bar Z-6H\vpbd\bo\bo'+12\vpbd^4\frac{\bar\alpha}{\Lambda^4}\big)\phi \nn\\
&&+ \frac{\bo'}{\bo}\pt_i^2\dvp + \frac{a^2}{2\bo^2}\vpbd\big(\bar Z+12\vpbd^2\frac{\bar\alpha}{\Lambda^4}\big)\dvpd 
- \frac{a^2}{2\bo^2}\big\{ \vpbdd\bar Z - 3(2\dot H+H^2)\bo\bo' + 12\vpbd^2\vpbdd\frac{\bar\alpha}{\Lambda^4} \big\} \dvp \nn\\
&&+ a\Big( H + \vpbd\frac{\bo'}{\bo} \Big)(\pt_i^2B-a\pt_i^2\dot E) \,, \\
\label{eqn:boxphi}
\pt_i^2\phi &=& \frac{a^2}{2\bo^2}\delta\rho - \frac{9a^2}{4\bo^2}H(\bar\rho+\bar p)\delta u - \frac{a^2}{\bo^2}\pt_i^2\pi 
+3a^2\Big(\vpbd\frac{\bo'}{\bo}-\frac{H}{2}\Big)\dot\psi - \frac{a^2}{2\bo^2}\big(\vpbd^2\bar Z - 3H(\vpbd\bo\bo'-H\bo^2) + 12\vpbd^4\frac{\bar\alpha}{\Lambda^4}\big)\phi \nn\\
&&- \frac{\bo'}{\bo}\pt_i^2\dvp + \frac{a^2}{2\bo^2}\vpbd\big(\bar Z+12\vpbd^2\frac{\bar\alpha}{\Lambda^4}\big)\dvpd 
- \frac{a^2}{4\bo^2}\big\{ (2\vpbdd-3H\vpbd)\bar Z - 3(2\dot H-5H^2)\bo\bo' + 12\vpbd^2(2\vpbdd-H\vpbd)\frac{\bar\alpha}{\Lambda^4} \big\} \dvp \nn\\
&&-H a(\pt_i^2B-2a\pt_i^2\dot E)- \vpbd\frac{\bo'}{\bo}a(\pt_i^2B-a\pt_i^2\dot E) -a\pt_i^2B+a^2\pt_i^2\ddot E \,, 
\eea}
the latter of which can be used to give the effective Newton constant. 
The anisotropic stress follows from the $i \neq j$ component of (\ref{eqn:1stij}).  Working in Longitudinal gauge with $E=B=0$ and for ordinary matter $\pi=0$ we find
{\small
\bea\label{eqn:sttas}
\psi - \phi = 2\frac{\Omega^\prime}{\Omega} \delta \varphi 
\eea}
We note that there is no anisotropic stress in the GR limit where $\bo=\;$1.

\section{$f(R)$ Models as Effective Field Theories \label{f(R)appendix}}
Next we turn to a brief discussion of $f(R)$ theories as effective field theories.  While this class of theories seem to belong to the class described by our master action Eq.~(1), we will see that a systematic expansion of the potential does not give rise to a valid EFT for cosmic acceleration today. In this Appendix only, we rescale the scalar $\vp$ to be dimensionless.

We reviewed in Sec.~\ref{sectionII}-2 that $f(R)$ is a scalar tensor theory with
\bea \label{fRparams}
\Omega^2=1+ f'(\vp)\,,\quad Z=0\,, \quad U=\frac{m_p^4}{2} \left(\vp f' - f \right)\,.
\eea
and $R= m_p^2\vp$. 

Let us next look at a general form for $f(R)$:
\begin{equation}
f(\vp) = c_n\vp^n.
\end{equation}
The potential is then \cite{faulkner}
\be
\tilde{U}(\vp) = \frac{m_p^4}{2} c_n(n-1) \Big( \frac{e^{\sqrt{\frac{2}{3}}\chi} - 1}{nc_n} \Big)^{n/(n-1)}e^{-2\sqrt{\frac{2}{3}}\chi}\,,
\ee
where we have made the definition $e^{\sqrt{\frac{2}{3}}\chi}=1+f'$. When $\chi \ll 1$, the scalar potential can be treated as an effective field theory.  Carrying out the expansion in the leading power of $\chi$, we find the scalar potential
\be
\tilde{U} \simeq \mu^4 \left(\sqrt{\frac{2}{3}}\,\chi \right)^{n/(n-1)}\,,
\ee
where 
\be
\mu^4 = \frac{c_n(n-1)}{2(nc_n)^{n/(n-1)}}m_p^4\,.
\ee
We immediately see that a sensible EFT for the scalar is obtained at the level of the effective potential only for $n > 1$ or $n < 0$.  To determine whether suitably sub-Planckian vevs can also be obtained, we derive the evolution of the scalar as a function of time, by including the coupling of the scalar field to matter:
\be
\tilde{U} \simeq \mu^4 \left(\sqrt{\frac{2}{3}}\chi\right)^{n/(n-1)} - \sqrt{\frac{2}{3}} \chi \rho_m \,.
\ee
Note that except for special points, neither $n > 1$ nor $n < 0$ give rise to stable potentials in general, with the contribution from the matter coupling balancing against the ordinary potential.\footnote{An exception is $n = 2$, which gives rise to a stable potential.}  Dark energy behavior in these potentials must arise from a Quintessence-like slow roll at late times. 

The equation of state is thus
\be
w+ 1 = \frac{2 T}{T + V} = \frac{\dot{\chi}^2}{3 \tilde{H}^2},
\ee
and using the equation of motion $\dot\chi=\tilde U'/3\tilde H$, we get
\be
w+1 = \frac{n^2 \tilde{U}^2}{3(n-1)^2 \rho_{tot}^2 \chi^2}.
\ee
Since $\tilde{U} \simeq \rho_{tot}$ in the accelerating epoch, in order to achieve the observational requirement of $w \approx -1$ we require {\em super}-Planckian vevs ($\chi \gg 1$), implying an inconsistency in this theory as an EFT for dark energy.\footnote{The models with potentials with $0 < n < 1$, although not making sense in terms of  an EFT expansion in powers of $\chi/m_p$, have stabilized minima. But the equation of state for $\chi$ is
\be
w+1 = - \frac{ \partial \log\tilde U}{3 \partial \log a} \,.
\ee
Since the potential has the minimum of
\be
\langle\tilde U\rangle = -\left(\frac{n-1}{\mu^4}\right)^{n-1}\left(\frac{\rho_m}{n}\right)^n\,,
\ee
we get
\be
w+1 = -\frac{a}{3\langle\tilde U\rangle}\frac{\d\langle\tilde U\rangle}{\d a} 
= - \frac{n}{3}\frac{a}{\rho_m}\frac{\d\rho_m}{\d a} = \frac{ns}{3}\,,
\ee
where we use $\rho_m \propto a^{-s}$. Thus, significant deviations from $w \approx -1$ does occur in this case too.}

\section{Comparison with the EFT of Quintessence \label{appendixg}}
Our master action (\ref{masteraction}) can reproduce generic Quintessence type models with EFT corrections: 
\be \label{kessmasteraction}
S = \int\d^4 x\sqrt{-g} \big\{ \frac{m_p^2}{2}R + \cL(\vp) \big\}\,,
\ee
where we ignore the matter contribution in order to concentrate on the scalar degrees of freedom, and we take the background
\be
\cL(\vp) = P(\vp,X) = \frac{1}{2}Z(\vp)X - U(\vp) + \frac{\alpha(\vp)}{\Lambda^4}X^2+ \cdots \,,
\ee
with $X = -g^{\mu\nu}\pt_\mu\vp\pt_\nu\vp$ and $\cdots$ representing terms higher in the derivative expansion.
Again, in this paper we will focus only on those backgrounds that admit such an expansion dropping higher order terms since they will be further suppressed by the cutoff.
The expansion of the action with respect to the metric and $\vp$ perturbations, (\ref{pertmetric}) and (\ref{perturbations}) respectively, can be obtained by straightforward algebra. Imposing the background EOMs
\bea
&&H^2 = \frac{1}{3m_p^2} \Big( -\bar\cL + 2\frac{\delta{\mc L}}{\delta g^{00}} \Big) = \frac{1}{3m_p^2} \big( -\bar P + 2P_X\vpbd^2 \big)\,, \\
&&\dot H + \frac{3}{2}H^2 = -\frac{1}{2m_p^2} \bar\cL = -\frac{1}{2m_p^2} \bar P \,, \\
\label{eqn:ke0thse}
&&P_\vp -2\pt_t(P_X\vpbd)-6HP_X\vpbd = 0\,,
\eea
we get
\bea\label{eqn:ourkep}
S_{\rm k-essence} &=& \frac{m_p^2}{2}\int \d^4 x\,a^3\,\Big[ \frac{2}{a^2}(\pt_i\psi)^2-\frac{4}{a^2}\pt_i\phi\pt_i\psi-6\dot\psi^2-12H\phi\dot\psi
-2\Big(\dot H+3H^2-\frac{2}{m_p^2}\frac{\delta^2{\mc L}}{(\delta g^{00})^2} \Big)\phi^2 \nn\\
&&\qquad -4\big(\frac{\pt_i^2B}{a}- \pt_i^2\dot E\big)(\dot\psi+H\phi) \Big] \nn\\
&&+\frac{1}{2}\int \d^4 x\,a^3\Big[ \frac{\delta^2{\mc L}}{\delta\dot\vp^2}\dot\dvp^2+\frac{\delta^2{\mc L}}{\delta\pt_i\vp\delta\pt_j\vp}\pt_i\dvp\pt_j\dvp
+\Big(\frac{\delta^2{\mc L}}{\dvp^2}-3H\frac{\delta^2{\mc L}}{\dvp\delta\dot\vp}-\pt_t\big(\frac{\delta^2{\mc L}}{\delta\vp\delta\dot\vp}\big) \Big) \delta\vp^2 \nn\\
&&\qquad +\Big( -4\frac{\delta^2{\mc L}}{\delta g^{00}\dvp}\phi + 12H\frac{\delta^2{\mc L}}{\delta g^{00}\delta\dot\vp}\phi 
+ 4\pt_t\big(\frac{\delta^2{\mc L}}{\delta g^{00}\delta\dot\vp} \big) \phi + 4\frac{\delta^2{\mc L}}{\delta g^{00}\delta\dot\vp} \dot\phi \nn\\
&&\qquad\qquad -2\frac{\delta{\mc L}}{\delta\dot\vp}(\dot\phi-3\dot\psi) + 4\frac{\delta^2{\mc L}}{\delta g^{0i}\delta\pt_j\vp}\frac{\pt_i\pt_j B}{a} -2\frac{\delta{\mc L}}{\delta\dot\vp}\pt_i^2 \dot E \Big)\delta\vp \Big] \nn\\
&=& \frac{m_p^2}{2}\int \d^4 x\,a^3\,\Big[ \frac{2}{a^2}(\pt_i\psi)^2-\frac{4}{a^2}\pt_i\phi\pt_i\psi-6\dot\psi^2-12H\phi\dot\psi
-2\Big(\dot H+3H^2-\frac{2}{m_p^2}P_{XX}\vpbd^4 \Big)\phi^2 \nn\\
&&\qquad -4\big(\frac{\pt_i^2B}{a}- \pt_i^2\dot E\big)(\dot\psi+H\phi) \Big] \\
&&+\frac{1}{2}\int \d^4 x\,a^3\Big[\, (4P_{XX}\vpbd^2 + 2P_X)\dot\dvp^2-\frac{2}{a^2}P_X(\pt_i\dvp)^2
+\big(P_{\vp\vp}-6HP_{X\vp}\vpbd-2\pt_t(P_{X\vp}\vpbd) \big) \delta\vp^2 \nn\\
&&\qquad +4\Big\{\big(-P_{X\vp}\vpbd^2+6H(P_{XX}\vpbd^3+P_X\vpbd)+2\pt_t(P_{XX}\vpbd^3+P_X\vpbd) \big)\phi \nn\\ 
&&\qquad\qquad +(2P_{XX}\vpbd^3+P_X\vpbd)\dot\phi + 3P_X\vpbd\dot\psi 
+P_X\vpbd \big(\frac{\pt_i^2B}{a}- \pt_i^2\dot E\big) \Big\} \dvp \,\Big]\,, \nn
\eea
where the variations of $\cL$ are evaluated at the background and $P_X=\frac{\partial P}{\partial X}\big|_0$, etc. We ignore the irrelevant zero-th order pieces, whereas the first order terms are eliminated by the background EOM's.

We can compare this with the literature. For example, \cite{Creminelli:2008wc} used $\cL(\vp) = P(X)$ and perturbed the scalar by
\be
\vp = \bar\vp + \vpbd \pi + \frac{1}{2}\vpbdd \pi^2 + \cdots\,,
\ee
to obtain the scalar action quadratic in $\pi$
\be\label{eqn:litkep}
S_{\pi,\,{\rm literature}} = \int\d^4 x\,a^3 \Big[\, (2P_{XX}\vpbd^4+P_X\vpbd^2)\dot\pi^2 - P_X\vpbd^2\frac{(\pt_i\pi)^2}{a^2} + 3\dot H P_X\vpbd^2\pi^2 \,\Big]\,.
\ee
On the other hand, with $P = P(X)$ and the obvious identification of
\be
\dvp = \vpbd \pi + \frac{1}{2}\vpbdd \pi^2 + \cdots\,,
\ee
the ${\mathcal O}(\pi^2)$ part of our (\ref{eqn:ourkep}) is
\bea
S_{\pi^2} &=& \int \d^4 x\,a^3\Big[\, (2P_{XX}\vpbd^2 + P_X)(\vpbd\dot\pi+\vpbdd\pi)^2-P_X\vpbd^2\frac{(\pt_i\pi)^2}{a^2} \,\Big] \nn\\
&=& \int \d^4 x\,a^3\Big[\, (2P_{XX}\vpbd^4 + P_X\vpbd^2)\dot\pi^2-P_X\vpbd^2\frac{(\pt_i\pi)^2}{a^2} 
-(3HQ\vpbd\vpbdd + \dot Q\vpbd\vpbdd + Q\vpbd\dddot{\bar\vp})\pi^2 \,\Big] \,,\qquad
\eea
where $Q = 2P_{XX}\vpbd^2+P_X$. Taking the time derivative of (\ref{eqn:ke0thse})
\be
0=\pt_t(2P_{XX}\vpbd^2\vpbdd+P_X\vpbdd+3HP_X\vpbd)
= \dot Q \vpbdd + Q \dddot{\bar\vp} + 3HQ\vpbdd + 3\dot HP_X\vpbd \,,
\ee
we get
\be
S_{\pi^2} = \int \d^4 x\,a^3\Big[\, (2P_{XX}\vpbd^4 + P_X\vpbd^2)\dot\pi^2-P_X\vpbd^2\frac{(\pt_i\pi)^2}{a^2} 
+ 3\dot HP_X\vpbd^2\pi^2 \,\Big] \,,
\ee
which agrees with the result appearing in (\ref{eqn:litkep}).

\section{Conformal Transformations \label{appendixc}}
We take the action in the Jordan frame,
\be 
S=  \int \d^Dx \sqrt{-g} \Big\{\frac{m_p^2}{2} F(\varphi) R - \half h(\varphi) \left( \partial \varphi \right)^2 - U(\varphi) +  \cL_m(\psi_m,g^{\mu \nu}) \Big\}\,,
\ee
and transform it to the Einstein frame by introducing a new metric
\be
\tilde{g}_{\mu \nu}=\Omega^2 g_{\mu \nu},
\ee
which depends explicitly on $\varphi$ -- demonstrating that in the new frame matter couplings depend on the evolution of the scalar and the motion of particles does not follow geodesics. 
Under this transformation we find
\bea
\sqrt{-g} &=& \Omega^{-D} \sqrt{-\tilde{g}}\,, \\
R&=&\Omega^2 \Big\{ \tilde{R} + 2(D-1) \tilde{\Box} \ln \Omega - (D-2)(D-1) \tilde{g}^{\mu \nu} \frac{\tilde{\partial}_\mu \Omega \tilde{\partial}_\nu \Omega}{\Omega^2} \Big\}\,,
\eea
so that the Einstein frame action becomes
\bea
\tilde{S}&=&  \int \d^Dx \sqrt{-\tilde{g}} \, \Omega^{-D}\left\{ \frac{m_p^2}{2}
\Omega^{2} F(\varphi) \Big(   \tilde{R} + 2(D-1) \tilde{\Box} \ln \Omega - (D-2)(D-1) \tilde{g}^{\mu \nu} \frac{\tilde{\partial}_\mu \Omega \tilde{\partial}_\nu \Omega}{\Omega^2} \Big) \right. \nonumber \\
 &&\hspace{80pt} \left. - \half h(\varphi) \Omega^{2} \big( \tilde{\partial}_\mu \varphi \big)^2 - U(\varphi) 
+ \cL_m(\psi_m,\Omega^2\tilde g^{\mu \nu}) \right\}.
\eea 
We see a canonical gravity term is possible by choosing $\Omega^2=F^{2/(D-2)}$.

Specializing to $D=4$ the Einstein frame quantities are related to the Jordan frame ones by
\bea
\sqrt{-\tilde{g}}&=& \Omega^4 \sqrt{-g}, \nonumber \\
\tilde{\Box} \sigma&=& \Omega^{-2} \left( \Box\sigma + 2 g^{\mu \nu} \Omega^{-1} \partial_\mu \Omega \partial_\nu \sigma  \right), \nonumber \\
\tilde{R}&=&\Omega^{-2} \left( R - 6\Box \ln \Omega -6 \frac{(\partial \Omega)^2}{\Omega^2} \right).
\eea
and with $\Omega^2=F$ we find
\be
\tilde{S} = \int \d^4x \sqrt{-\tilde{g}} \left\{ \frac{m_p^2}{2} \tilde{R} - \half\, \frac{3m_p^2{F^\prime}^2 +2 h F}{2F^2} \big( \tilde{\partial}_\mu \varphi \big)^2 - F^{-2} U(\varphi) +   F^{-2}\cL_m(\psi_m,\Omega^2\tilde{g}^{\mu \nu}) \right\}, \label{eframe} 
\ee
where a prime denotes the derivative with respect to $\varphi$.

\end{document}